\documentclass[lettersize,journal]{IEEEtran}
\usepackage{aliascnt}
\usepackage{cite}
\usepackage{setspace}
\usepackage{wrapfig}
\usepackage{gensymb}
\usepackage{soul}

\IEEEoverridecommandlockouts
\newenvironment{talign}
 {\align}
 {\endalign}
\usepackage{multirow}
\usepackage{enumitem}
\usepackage{pifont}

\usepackage{amsmath,amssymb,amsfonts,amsthm}
\usepackage{algorithmic}
\usepackage[ruled,linesnumbered]{algorithm2e}
\usepackage{graphicx}
\usepackage{textcomp}
\usepackage{caption}
\usepackage {epstopdf}
\usepackage[hidelinks]{hyperref}
\usepackage{float}
\usepackage[utf8]{inputenc}
\usepackage{tabularx}
\usepackage{booktabs}
\usepackage{cases}
\usepackage{xcolor}
\usepackage{authblk}
\usepackage{cleveref}
\usepackage{subfigure}
\usepackage{subcaption}
\usepackage{bm}
\def\BibTeX{{\rm B\kern-.05em{\sc i\kern-.025em b}\kern-.08em
    T\kern-.1667em\lower.7ex\hbox{E}\kern-.125emX}}

\usepackage{graphicx}
\usepackage{mathrsfs}

\usepackage{bm}
\usepackage{hyperref,bookmark}

\DeclareRobustCommand{\vect}[1]{\bm{#1}}
\newtheoremstyle{slanted}
{0em plus 0em minus 0em}
  {0em plus 0em minus 0em}
  {\em}
  {}
  {\bfseries}
  {.}
  { }
  {}

\theoremstyle{slanted}

\theoremstyle{slanted}

\theoremstyle{slanted}

\theoremstyle{slanted}

\theoremstyle{slanted}

\theoremstyle{slanted}

\theoremstyle{slanted}

\usepackage{enumitem}

\title{User Connection and Resource Allocation Optimization in Blockchain Empowered Metaverse over 6G Wireless Communications} 
\author{Liangxin~Qian,~\IEEEmembership{Graduate Student Member, IEEE},
        Chang~Liu,~Jun~Zhao,~\IEEEmembership{Member, IEEE}
\thanks{L. Qian, C. Liu, and J. Zhao are with the School of Computer Science and Engineering at Nanyang Technological University (NTU), Singapore. (Emails: qian0080@e.ntu.edu.sg, liuc0063@e.ntu.edu.sg, and junzhao@ntu.edu.sg). \newline \indent
A 7-page shorter conference version \cite{qian2023user} is accepted for VTC 2024 Spring. According to the policy of IEEE ComSoc at \href{https://www.comsoc.org/publications/journals/ieee-transactions-wireless-communications/conference-vs-journal}{https://www.comsoc.org/publications/journals/ieee-transactions-wireless-communications/conference-vs-journal}, the journal and conference manuscripts can be submitted at the same time. Differences between the two versions are discussed in detail in this journal submission.
}}

\begin{document}
\maketitle

\begin{abstract}
The convergence of blockchain, Metaverse, and non-fungible tokens (NFTs) brings transformative digital opportunities alongside challenges like privacy and resource management. Addressing these, we focus on optimizing user connectivity and resource allocation in an NFT-centric and blockchain-enabled Metaverse in this paper. Through user work-offloading, we optimize data tasks, user connection parameters, and server computing frequency division. In the resource allocation phase, we optimize communication-computation resource distributions, including bandwidth, transmit power, and computing frequency. We introduce the trust-cost ratio (TCR), a pivotal measure combining trust scores from users' resources and server history with delay and energy costs. This balance ensures sustained user engagement and trust. The DASHF algorithm, central to our approach, encapsulates the \underline{D}inkelbach algorithm, \underline{a}lternating optimization, \underline{s}emidefinite relaxation (SDR), the \underline{H}ungarian method, and a novel \underline{f}ractional programming technique from a recent IEEE JSAC paper~\cite{zhao2023human}. The most challenging part of DASHF is to rewrite an optimization problem as Quadratically Constrained Quadratic Programming (QCQP) via carefully designed transformations, in order to be solved by SDR and the Hungarian algorithm. Extensive simulations validate the DASHF algorithm's efficacy, revealing critical insights for enhancing blockchain-Metaverse applications, especially with NFTs.

\end{abstract}

\setlength{\abovedisplayskip}{0pt plus 0pt minus 0pt}
\setlength{\belowdisplayskip}{0pt plus 0pt minus 0pt}
\setlength\abovedisplayshortskip{0pt plus 0pt minus 0pt}
\setlength\belowdisplayshortskip{0pt plus 0pt minus 0pt}

\section{Introduction}
The connection of blockchain technology, Metaverse platforms, and seamless connectivity of non-fungible token (NFT) applications signifies a remarkable milestone in the further development of wireless communication systems \cite{wang2022toward,lim2022realizing,xu2022full,xu2023epvisa}. The Metaverse, as a burgeoning immersive digital space, offers a platform for users to delve into meticulously designed virtual universes. Integral to this virtual realm, NFTs emerge to provide users with unique ownership over digital items and experiences \cite{wu2023virtual}. 
Making sure NFTs work well in the Metaverse relies on the strength of 6G wireless communication systems. These systems need to provide continuous, high-quality user experiences without using too much energy or causing delays \cite{wang2022survey,christodoulou2022nfts}.


In this digital world, blockchain serves as the foundation, giving the Metaverse trust, security, and transparency. It secures NFT transactions and gives users proof of ownership of their digital assets~\cite{chalmers2022beyond}. But, setting up user connections and managing resources in a Metaverse that uses blockchain reveals many complex challenges \cite{cheng2022will,aggarwal2019blockchain,lu2019blockchain}.

This paper delves into the collaborative optimization of user connections and resource allocation to augment communication and computation efficiencies in the blockchain-infused Metaverse, particularly within the NFT context. As we navigate through our research, we underscore the paramount challenges and motivations that shape our journey.

\textbf{Challenges and Motivations.} The rapid growth of the Metaverse and its integration with blockchain technology introduces a range of challenges and opportunities, especially when considering wireless communications. The necessity for enhanced privacy and efficient resource allocation becomes apparent as the Metaverse's complex web of user interactions requires secure spaces for digital identities and transactions \cite{yang2022fusing,huang2023security,di2021metaverse}. Blockchain’s core features offer promising solutions for improving privacy and security. However, merging blockchain into the Metaverse introduces complexities \cite{kang2023security,far2022applying}, highlighting the need for creative resource management strategies. Prior works, such as those by Feng \emph{et~al.} \cite{feng2020joint}, have made progress in optimizing connections and resources. Yet, their approach to simplifying bandwidth and power allocation points to a potential oversight in fully leveraging resource optimization. This trend of oversight is further evident in the work of Dai \emph{et~al.} \cite{dai2018joint}. Their research, which seeks to optimize computation offloading and user association, does break new ground. However, the concentration on computation frequency and transmit power as the main resources fails to capture the broader spectrum of challenges. This situation emphasizes the importance of a comprehensive approach that considers the full spectrum of challenges, including bandwidth, latency, energy consumption, reliability, as well as computational resource use, transmit power use, and network design.

The integration of blockchain within the wireless-enabled Metaverse also brings to light essential prerequisites and significant challenges, focusing on work offloading with higher trust, energy consumption, and latency. An efficient blockchain architecture is vital to support the high transaction volumes and real-time interactions of the Metaverse. Energy consumption emerges as a key challenge, with traditional blockchain models being notably high in energy usage. Finding energy-efficient blockchain solutions is critical for the Metaverse's sustainability and for providing a smooth user experience. Additionally, reducing latency for real-time activities and effectively managing user-server connection and work offloading ratio with higher trust underscores the need for solutions that tackle both technological and operational challenges in this emerging space.

Our research is driven by the imperative to address these complexities, aiming to fill the gaps in combining blockchain with the Metaverse over wireless networks. We are motivated to reduce latency and energy consumption, ensure user-server connection and work offloading ratio with higher trust, and improve communication and computational resource allocation.

\textbf{Studied Problem.}
In this paper, we study the blockchain-empowered Metaverse system and introduce the pioneering concept of trust-cost ratio (TCR), which is a ratio of trust score and cost consumption ($\frac{\text{trust score}}{\text{cost}}$). This value is an ingenious metric that encapsulates the delicate equilibrium between user trust scores and the overarching considerations of delay and energy consumption across the communication spectrum. The trust score includes a user's radio and computing resources from the server and a score based on the server's historical experience with data processing and blockchain work. Cost consumption includes the maximum delay and total energy consumption of the users and servers. We assume that the user does partial work offloading first; that is, the user offloads part of the work to the server, and the rest of the work is handled locally. On the server side, the server first receives the work offloaded from the user and then allocates radio and computing resources for each user, including bandwidth, transmit power of users and servers, and computing frequency of users and servers. In addition, the server further divides the computing resources allocated to each user. Specifically, part of the computing resources are used to process the data processing work of the user offload, and the rest is used to generate, upload, and verify the blockchain.

\subsection{Related Work}
We discuss the related work in the following parts: resource allocation in blockchain-based wireless networks, trust model in blockchain, and blockchain with Metaverse.

\textbf{Resource allocation in blockchain-based wireless networks.} In the context of blockchain-based wireless networks, several notable works have contributed to the optimization of resource allocation and performance enhancement. Feng \emph{et~al.} \cite{feng2020joint} delve into the joint optimization of user connections, radio resources, and computational resource allocation within a blockchain-based mobile edge computing (MEC) system~\cite{lv2021strategy}. 
It's noteworthy that their treatment of data rate as a direct variable may not offer a precise delineation for bandwidth and transmit power allocation, as these vital resource aspects lack explicit detailing. Furthermore, their optimization objectives primarily center around minimizing energy consumption and reducing delays, leaving other critical resource aspects unexplored.
Similarly, Dai \emph{et~al.} \cite{dai2018joint} propose a framework addressing the joint optimization of computation offloading and user association in MEC systems. Yet, it's important to highlight that their optimization efforts primarily target computation frequency and transmit power as the key resources to optimize. Additionally, their optimization problem primarily focuses on minimizing energy consumption without addressing communication resource allocation or adopting a holistic approach that encompasses user association within a blockchain-enabled Metaverse. Guo \emph{et~al.} \cite{guo2019adaptive} present a comprehensive blockchain-based MEC framework, integrating deep reinforcement learning~\cite{li2024trajectory} to achieve consensus among nodes while optimizing both MEC and blockchain systems. Their focus is on adaptive resource allocation, block size management, and block generation to enhance the system throughput and ensure superior quality of service (QoS) for users in future wireless networks. Sun \emph{et~al.} \cite{sun2020joint} introduce two double auction~\cite{zheng2014star,zheng2014unknown} mechanisms within a blockchain-driven MEC framework, aiming to ensure reliable cross-server resource allocation and protect against tampering of user information by malicious edge servers. This approach leads to improved MEC system efficiency. 
Additionally, Xu \emph{et~al.} \cite{xu2021edge} propose a Stackelberg dynamic game-based resource pricing and trading scheme seamlessly integrated with blockchain technology. This scheme optimizes allocating edge computing resources between edge computing stations (ECSs) and unmanned aerial vehicles (UAVs) in mobile networks while addressing security and privacy concerns, validated through numerical simulations.
Furthermore, Jiang \emph{et~al.} \cite{jiang2020intelligent} introduce a video analytics framework incorporating multiaccess edge computing and blockchain technologies into vehicular networks. Their work addresses challenges related to data transmission and security, demonstrating the efficacy of their approach in optimizing blockchain transaction throughput and reducing MEC latency through deep reinforcement learning. Collectively, these research efforts provide a comprehensive exploration of resource allocation, consensus mechanisms, and performance optimization in blockchain-based wireless networks.

\textbf{Trust model in blockchain.} In the realm of blockchain-based trust models, several significant contributions have emerged, each addressing distinct aspects of trust and security in diverse IoT scenarios. Tu \emph{et~al.} \cite{tu2022blockchain} introduce a blockchain-based trust and reputation model (BTRM) geared towards fortifying IoT security by conducting a comprehensive evaluation of user reputation and resistance against multiple malicious attacks. Complementing this, they present a dynamic evaluation mechanism (DEM) within a hyper ledger fabric prototype system, which streamlines reputation evaluations while preserving network security. Liu \emph{et~al.} \cite{liu2022semi} propose a semi-centralized trust management system based on blockchain, tailored to facilitate IoT data exchange. Their innovative computational trust model incorporates decay functions and adaptable weights, effectively identifying and mitigating the influence of malicious devices, as substantiated through simulation experiments, positioning it favorably against conventional models.
Finally, Xi \emph{et~al.} \cite{xi2023blockchain} present a dynamic blockchain sharding scheme founded on the hidden Markov model (HMM), which efficiently reduces cross-shard transactions while concurrently boosting system throughput and reducing transaction confirmation latency, especially valuable for collaborative IoT applications. 
To sum up, these pioneering works offer a diverse spectrum of trust models, each tailored to bolster security and trustworthiness in various IoT contexts.
\textbf{Blockchain with Metaverse.} In the domain of blockchain's integration with the Metaverse, two noteworthy studies have contributed to addressing distinct challenges and advancing the capabilities of this convergence. Hoa \emph{et~al.} \cite{hoa2023dynamic} consider an edge computing-assisted Metaverse system featuring a virtual service provider (VSP) entrusted with offloading data collected by UAVs for updating digital twins (DTs). This research strategically confronts dynamic data management, user latency requirements, and resource constraints, employing a stochastic problem formulation and leveraging deep reinforcement learning. 
Together, the above studies showcase the evolving landscape of blockchain and Metaverse integration, addressing specific issues while harnessing the potential of cutting-edge technologies to shape the future of digital ecosystems.

\textbf{Differences with the conference version \cite{qian2023user}.} In \cite{qian2023user}, the focus is on collaborative server training involving neural network adapters within the context of large language model-based applications to mitigate computational and communication overhead, thereby guaranteeing that users have access to sufficient computational and communication resources. In our work, however, we shift the spotlight to efficient offloading tasks and enhancing user-server connections within the burgeoning domain of a blockchain-powered Metaverse, specifically targeting NFT applications. This pivot necessitates a fresh approach to the communication and computation formulas applied.
Whereas \cite{qian2023user} explored the trade-off between user experience scores and total cost, our investigation pivots towards balancing user trust scores against total cost. Our contribution extends the groundwork laid by \cite{qian2023user} by incorporating a model that captures user trust in servers tasked with processing blockchain operations. While \cite{qian2023user} was limited to optimizing resource allocation alongside user connection and training ratio, our paper delves further, tackling the nuanced optimization of user connection, offloading ratio, and the division of server computing frequencies.
The methodology in \cite{qian2023user} doesn't specify an algorithmic procedure, whereas our paper delineates three intricate algorithmic methodologies. Building upon the comparative baseline established by \cite{qian2023user}, our paper enriches the discourse with simulation results across diverse server computation frequencies and offers an expanded suite of figures demonstrating algorithm convergence in various system configurations.

\subsection{Main Contributions}
Our contributions include a novel joint optimization problem, the introduction of the TCR metric, and an algorithm to solve the problem, detailed as follows:
\begin{itemize}
\item[$\bullet$]Introducing the Trust-Cost Ratio: We introduce the concept of the trust-cost ratio (TCR). To the best of our knowledge, there is no existing research studying this. TCR quantifies the balance between user trust scores and the overall delay and energy consumption in the entire uplink and downlink communications. It provides a valuable metric for assessing the trade-off between user trust and resource efficiency.
\item[$\bullet$]Apart from the traditional optimization of partial work offloading, communication resource allocation, and computation resource allocation of users and servers, we also consider the computation resource allocation of servers for different tasks (i.e., data processing tasks and blockchain tasks). This new optimization variable is unique in blockchain-based wireless communication systems.
\item[$\bullet$]An Algorithm to Solve the Proposed Optimization: To address the joint optimization problem discussed above, our developed algorithm, termed DASHF, leverages the combination of the \underline{D}inkelbach algorithm, \underline{a}lternating optimization, \underline{s}emidefinite relaxation (SDR), the \underline{H}ungarian algorithm, and a novel \underline{f}ractional programming (FP) technique by~\cite{zhao2023human}  published in IEEE JSAC recently. The most challenging part of DASHF is to rewrite an optimization problem as Quadratically Constrained Quadratic Programming (QCQP) via carefully constructed transformations to leverage SDR and the Hungarian algorithm to obtain a solution. In DASHF, we first tackle the optimization of user connections, work-offloading ratios, and server computing frequency division as QCQP. Subsequently, we delve into optimizing communication-computation resource allocation (for bandwidth, transmit power of users and servers
and computing frequency of users and servers), providing an effective solution for the challenging non-convex FP problem.
\item[$\bullet$]The simulation results confirm that the proposed DASHF algorithm effectively optimizes user work offloading, resource allocation, trust-cost ratio, and communication-computation resources.
\end{itemize}

The rest of this paper is organized as follows. Section~\ref{section.System Model} presents the system model and optimization formulation. We propose the DASHF algorithm to solve the optimization problem in Section \ref{section.proposed AO technique}, with its complexity discussed in Section~\ref{section.Complexity Analysis}. The numerical results are provided in Section~\ref{section.Numerical Results}. We conclude the paper in Section \ref{section.Conclusion}.
\begin{figure*}[t]
\vspace{-0.9cm}
\centering
\includegraphics[width=0.9\textwidth]{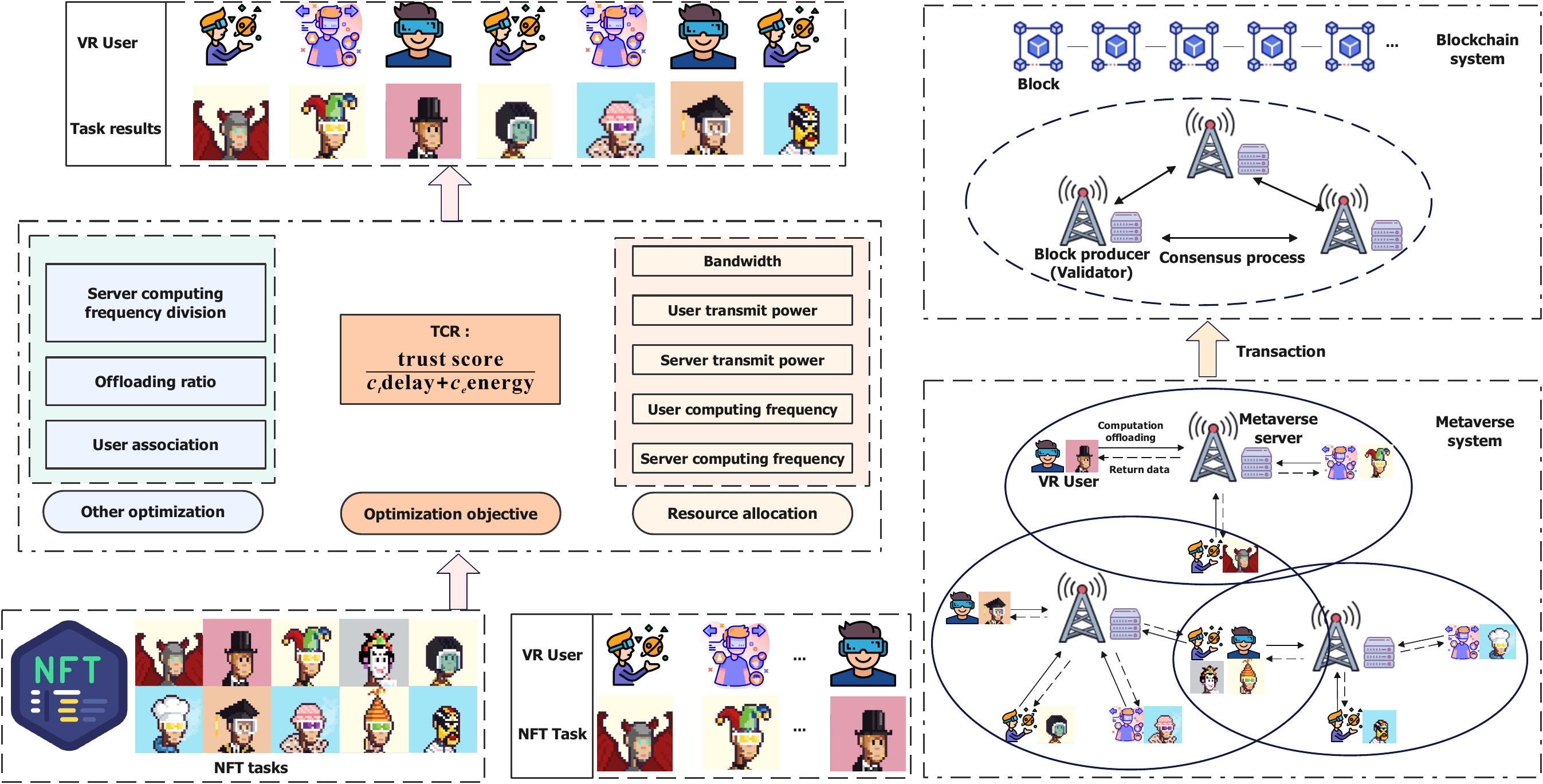}
\caption{Optimizing the trust-cost ratio (TCR) of a system consisting of $N$ VR users and $M$ Metaverse servers by the joint optimization of user association, offloading ratio, server computing capacity division, and radio and computing resource allocation. The images in the NFT tasks come from \href{https://monkeykingdom.io/}{Monkey Kingdom}\cite{monkeykingdom}.}\vspace{-5pt}
\label{Fig.system}
\end{figure*}
\section{Blockchain-Empowered Metaverse System and Optimization Problem Formulation}\label{section.System Model}
In this section, we will introduce the studied system scenario and formulate the optimization problem.

\subsection{An Overview of the System}
As presented in Fig. \ref{Fig.system} on the next page, we consider a blockchain-empowered Metaverse system that operates within the realm of wireless communication. Within this envisioned framework, each VR user is assigned a specific set of NFT tasks, necessitating partial delegation to the Metaverse server. This Metaverse server assists in data processing, as well as functioning as the generator and validator of blockchain transactions. Once the Metaverse server concludes the processing of user-generated data, it proceeds to upload computation offloading records onto the blockchain system, where they are systematically documented within the distributed ledger. These records are categorized as individual ``transactions" within the blockchain system. Subsequently, these transactions undergo meticulous verification by other Metaverse servers before being consolidated into a new block. The achievement of a consensus across the entirety of the Metaverse network heralds the seamless integration of the new block into the overarching blockchain structure. Following this integration, the Metaverse server takes on the responsibility of disbursing the processed data back to the respective users, ensuring an efficient data distribution mechanism within the system.
\subsection{The Detailed System Model}

We consider a system comprising $N$ VR users and $M$ Metaverse servers. We use $n$ and $m$ as indices for a VR user and a Metaverse server, respectively, where $n \in \mathcal{N} := \{1,2,\cdots,N\}$ and $m \in \mathcal{M} := \{1,2,\cdots,M\}$. Each user is connected to one and only one server; i.e., $\sum_{m\in \mathcal{M}} x_{n,m} = 1$. We introduce indicator variables $x_{n,m}\in\{0,1\}$ to characterize the connection between users and servers; specifically, $x_{n,m}=1$ (resp, $0$) means that the \mbox{$n$-th} user is connected (resp., not connected) to the \mbox{$m$-th} server. For example, if $x_{n,m}=1$, it means that the \mbox{$n$-th} user only connects to the \mbox{$m$-th} server and $x_{n,m'}=0$ for $m{'}\in\mathcal{M} \setminus \{m\}$.

\subsubsection{Time consumption}\label{secTime}
We consider frequency-division multiple access (FDMA) so that communication among users and servers would not interfere. Let $b_{n,m}$ be the allocated bandwidth for the communication between user $n$ and server $m$, and $p_n$ be the transmit power of user $n$. Then we define the transmission rate from user $n$ to the chosen edge server $m$ as $ r_{n,m} = b_{n,m} \log_2(1+\frac{g_{n,m}p_{n}}{\sigma^2b_{n,m}})$ according to the \mbox{well-known} Shannon formula,
where $\sigma^2$ is the noise power spectral density, and $g_{n,m}$ is the channel attenuation. We further express $g_{n,m}$ as $g_{n,m} = h_{n,m}l_{n,m}$, with $h_{n,m}$ being the large-scale \mbox{slow-fading} component capturing effects of path loss and shadowing and $l_{n,m}$ being the small-scale Rayleigh fading.

\textbf{Work offloading and data processing phases.} The total task data size at user $n$ is $d_n$ bits. The offloading task data size is $\varphi_n d_n$ for $\varphi_n \in [0,1]$, and the rest of the task (i.e., the $(1-\varphi_n) d_n$ part) will be processed by the user itself. After the user-server connection algorithm (this can be completed by choosing the nearest neighbor server sets, then choosing the server with the lowest transmission time, and finally finishing all user-server connections). The transmission time from the user $n$ to the server $m$ is $T^{(t_1)}_{n,m} = \frac{x_{n,m}\varphi_n d_n}{r_{n,m}}$, where the superscript ``$t$'' represents transmission.
User $n$ processes $(1-\varphi_n) d_n$ by itself, and the processing time is $T^{(p_1)}_{n} = \frac{(1-\varphi_n)d_n f_n}{F_n}$, where the superscript ``$p$'' stands for processing,
 $f_n$ (cycles/bit) is the CPU clock cycles per bit required by the \mbox{$n$-th} user, and $F_n$ is the CPU cycle frequency of the \mbox{$n$-th} user. Below, we look into server computation. After using $F_{n,m}$ to denote the CPU cycle frequency of the \mbox{$m$-th} server allocated to \mbox{$n$-th} user, we let $\gamma_{n,m} \in (0,1)$ be the fraction of $F_{n,m}$ used to process user $n$'s data, and $(1-\gamma_{n,m})$ be the fraction of $F_{n,m}$ used for the blockchain task. Then for  server $m$ to process the  data of $\varphi_n d_n$ bits sent from user $n$, the processing delay is $T^{(p_1)}_{n,m} = \frac{x_{n,m}\varphi_n d_n f_m}{\gamma_{n,m} F_{n,m}}$,
 where $f_m$ is the CPU clock cycles per bit required by server $m$ and $x_{n,m}$ is just the binary indicator variable which has been defined at the beginning of this subsection. We consider $\sum\nolimits_{n \in \mathcal{N}} x_{n,m} F_{n,m} \leq F_m$, where $F_m$ is the maximum CPU cycle frequency of server $m$.

\textbf{Blockchain phase.} With the $\varphi_n d_n$ data bits sent from user $n$, server $m$ will generate a blockchain block for such data. For this block generation, we consider that the server will process $\varphi_n d_n \omega_b$ bits, where $\omega_b$ represents the data size change ratio when server $m$ maps user $n$'s data to a format that can be processed by the blockchain. Recall that we have explained in the previous paragraph that $(1-\gamma_{n,m})$ fraction of $F_{n,m}$ is used for the blockchain task. Hence, the delay for server $m$ to generate the block for user $n$ by leveraging the computing capability of $(1-\gamma_{n,m})F_{n,m}$ can be expressed by 
$T^{(g)}_{n,m} = \frac{x_{n,m}\varphi_n d_n \omega_b f_m}{(1-\gamma_{n,m})F_{n,m}}$, where the superscript ``$g$'' means generation, and $f_m$ has been defined in the previous paragraph.
Next, we discuss the block propagation. We consider only one hop among servers and utilize the result of \cite{feng2020joint} to obtain the following. The total block propagation time used in data transactions during the consensus is $T^{(bp)}_{n,m} = \frac{S_b}{R_{m}}$, where the superscript ``$bp$'' stands for block propagation,
and $R_{m} := \mathop{\min}\nolimits_{m{'} \in \mathcal{M} \setminus \{m\}} R_{m,m'}$, with $R_{m,m'}$ denoting the wired link transmission rate between servers $m$ and $m'$. Note that ``$\min$'' means that the block generated by server $m$ has to be propagated to all other $M-1$ servers (taking ``$\min$'' among the rates is like computing ``$\max$'' among the delays). Then, during the validation phase, all $M-1$ servers do the verification work. According to \cite{feng2020joint}, the validation time is $T_{n,m}^{(v)} = \mathop{\max}_{m{'} \in \mathcal{M} \setminus \{m\}} \frac{f_v}{(1-\gamma_{n,m'})F_{n,m'}}$,
where the superscript ``$v$'' is short for verification, and $f_v$ denotes the CPU cycles required by server $m{'}$ to verify the block.

\textbf{Returning data and data processing phases.} We use $\omega_p$ to denote the data size change ratio from the raw data to the processed data. Thus, the processed data size can be expressed as $\omega_p \varphi_n d_n $. Server $m$ transmits the results to the user $n$, and the delay is $T^{(t_2)}_{n,m} = \frac{x_{n,m}\omega_p \varphi_n d_n}{r_{m,n}}$,   
where the superscript ``$t$'' stands for transmission. We assume that the path loss and bandwidth between the downlink and uplink are the same.  Then, user $n$ processes the received data, and the delay is $T^{(p_2)}_n = \frac{\omega_p \varphi_n d_n f_n}{F_n}$, where the superscript ``$p$'' represents processing.
The time on the server side is
\begin{talign}
    T_{s,n,m} = T^{(t_1)}_{n,m} + T^{(p_1)}_{n,m} + T^{(g)}_{n,m} + T^{(bp)}_{n,m} + T_{n,m}^{(v)},
\end{talign}
where the subscript ``$s$'' represents ``server''.
The time consumed on the user side is 
\begin{talign}
    T_{u,n,m} = T^{(p_1)}_{n} + T^{(t_2)}_{n,m} + T^{(p_2)}_{n},
\end{talign}
where the subscript ``$u$'' stands for ``user''.
With the above notation, the system delay is computed as
\begin{talign}
    T_{\text{total}} = \max\limits_{n\in \mathcal{N}, m\in \mathcal{M}} \big\{\max\left\{ T_{s,n,m}, T_{u,n,m}\right\}\big\}.\label{eqTtotal}
\end{talign}

\begin{figure}[tbp]
\centering
\vspace{-15pt}\includegraphics[width=0.48\textwidth]{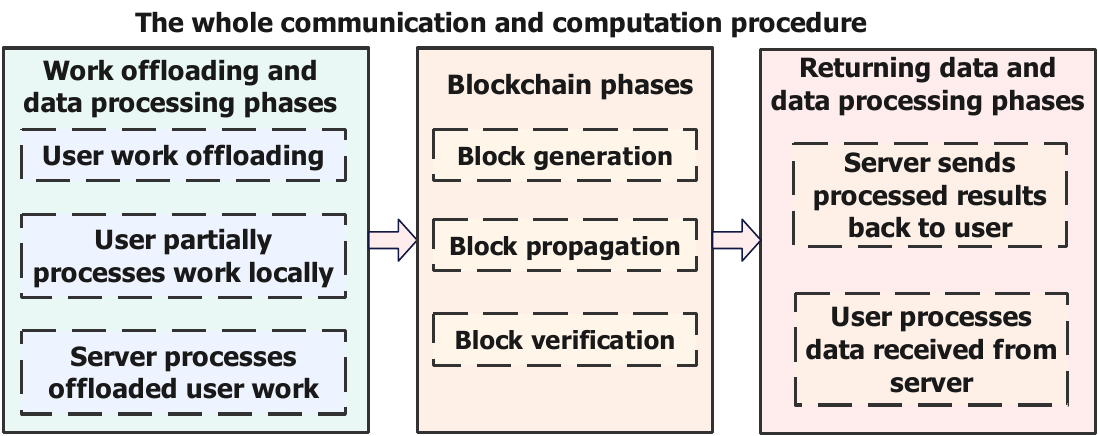}
\vspace{-2pt}\caption{The whole communication and computation procedure.}
\label{Fig.whole_procedure}
\end{figure}

\textbf{The whole communication and computation procedure.} In the proposed system, the procedure begins with users offloading computational tasks to servers while simultaneously processing part of the work locally. The server processes these offloaded tasks and generates transactions that are packaged into blocks. These blocks then undergo a consensus mechanism, involving propagation to and verification by other block producers, to be added to the blockchain. This ensures the integrity and security of the transactions related to the user's tasks. In the consensus, the delegated proof-of-stake (DPoS) consensus algorithm is considered and all validators are assumed to be honest \cite{feng2020joint}. Finally, the server returns the processed results to the user, who then performs any necessary further processing. Throughout this system, servers not only handle computation but also act as witnesses in recording transactions on the blockchain, facilitating a secure, efficient, and transparent workflow from task offloading to final data processing. The whole communication and computation procedure is also presented in Fig. \ref{Fig.whole_procedure}.

\subsubsection{Energy consumption} \label{secEnergy}
Based on the delay discussion, we then compute the energy consumption in this system. Energy used for transmitting data from the \mbox{$n$-th} user to the server $m$ is given as $ E^{(t_1)}_{n,m} = p_n T^{(t_1)}_{n,m} = p_n \frac{x_{n,m}\varphi_n d_n}{r_{n,m}}$.
Energy used for user $n$ processing data locally can be calculated as follows based on~\cite{yang2020energy}: $E^{(p_1)}_{n} = \kappa_n (1-\varphi_n) d_n f_n F^{2}_n$,
where $\kappa_n$ denotes the effective switched capacitance that depends on the chip architecture of user $n$. We omit the energy caused by the blockchain's propagation and validation process since they are negligible compared with those for other stages. In other words, we are mainly concerned with the energy caused by the server processing data, generating blocks, and transmitting data. Energy for server $m$ to process $\varphi_n d_n$ data is given as $E^{(p_1)}_{n,m} = \kappa_m x_{n,m} \varphi_n d_n f_m (\gamma_{n,m} F_{n,m})^{2}$,
where $\kappa_m$ denotes the effective switched capacitance that depends on the chip architecture of server $m$. Energy used for the server to generate one block is $E^{(g)}_{n,m} = \kappa_m x_{n,m} \varphi_n d_n \omega_b f_m [(1-\gamma_{n,m})F_{n,m}]^2$.
Energy caused by server $m$ to transmit the processed data to user $n$ is $E^{(t_2)}_{n,m} = p_m T^{(t_2)}_{n,m} = p_m \frac{x_{n,m} \varphi_n d_n \omega_p}{r_{m,n}}$.
Energy caused by user $n$ to process $\varphi_n d_n \omega_p$ bits of data is $E^{(p_2)}_n = \kappa_n \varphi_n d_n \omega_p f_n F_n^2$.
We denote the energy consumed at user side as $E_{u,n,m} = \sum_{n \in \mathcal{N}} \{E^{(p_1)}_{n} + E^{(p_2)}_{n}\} + \sum_{n\in \mathcal{N}, m\in \mathcal{M}} E^{(t_1)}_{n,m}$ and the energy consumed at server side as $E_{s,n,m} = \sum_{n\in \mathcal{N}, m\in \mathcal{M}} \{E^{(p_1)}_{n,m} + E^{(g)}_{n,m} + E^{(t_2)}_{n,m}\}$. Thus, the system (i.e., total) energy consumption can be formulated as
\begin{talign}
    E_{\text{total}} & \hspace{-2pt} =\hspace{-2pt}  \sum\limits_{n \in \mathcal{N},m\in \mathcal{M}}\hspace{-2pt} (E_{u,n,m} + E_{s,n,m}) \hspace{-2pt}  =\hspace{-2pt}  \sum\limits_{n \in \mathcal{N}} \{E^{(p_1)}_{n} 
+ E^{(p_2)}_{n}\}  \nonumber\\[-3pt]
    &+ \sum\limits_{n\in \mathcal{N}, m\in \mathcal{M}} \{E^{(t_1)}_{n,m} + E^{(p_1)}_{n,m} + E^{(g)}_{n,m} + E^{(t_2)}_{n,m}\}.\label{eqEtotal}
\end{talign}\vspace{-3pt}

\begin{table}[tbp]
\setlength{\abovecaptionskip}{5pt}
\setlength{\belowcaptionskip}{10pt}
\caption{Important Notation.}\vspace{-3pt}\label{table_notation}
\centering
\setlength{\tabcolsep}{0pt}
\small
\renewcommand{\arraystretch}{1.1}
\begin{tabular}{m{5em}  m{7cm}}
    \toprule                                   
    \textbf{Notation}  & \textbf{Description} \\
    \midrule
    $\mathcal{N}$ & The set of all users ($n \in\{1,...,N\}$)\\ 
    $\mathcal{M}$ & The set of all servers ($m \in\{1,...,M\}$)\\
    $x_{n,m}$ & The user connection between user $n$ and server~$m$\\
    $d_n$ & The total work of user $n$\\
    $p_{n}$ & The transmit power of user $n$\\
    $p_{max}^{(n)}$ & The maximum transmit power of user $n$\\
    $p_{n,m}$ & The transmit power of server $m$ allocated for user $n$\\
    $p_{max}^{(m)}$ & The maximum transmit power of server $m$\\
    $b_{n,m}, b_{m,n}$~ & The allocated bandwidth between user $n$ and server~$m$\\
    $b_{max}$ & The maximum allocated bandwidth for each server\\
    $g_{n,m}$ & The channel attenuation from user $n$ to server~$m$\\
    $r_{n,m}$ & The transmission rate from user $n$ to server $m$ \\
    $r_{m,n}$ & The transmission rate from server $m$ to user $n$ \\
    $v_{n,m}$ & The trust score value of user $n$ connected to server~$m$ \\
    $\varphi_n$ & The offloading work ratio of user $n$\\
    $F^{(n)}_{max}$ & The maximum CPU cycle frequency of user $n$\\
    $F^{(m)}_{max}$ & The maximum CPU cycle frequency of server $m$\\
    $f_n, f_m$ & The CPU clock cycles per bit of user $n$ and server~$m$\\
    $F_{n,m}$ & The total computational capacity of server $m$ allocated for user~$n$\\
    $\gamma_{n,m}$ & The computational capacity ratio of server $m$ allocated for processing data for user $n$\\
    $\omega_b$ & The data size change ratio between the MEC task and blockchain task\\
    $\omega_p$ & The data size change ratio between the raw data and processed data\\
    $\omega_t$ & The weight parameter of delay cost\\
    $\omega_e$ & The weight parameter of energy cost\\
    $S_b$ & The block size\\
    $R_{m}$ & The wired link transmission rate among the servers\\
    $\tau$ & The scoring parameter that is based on the server's history of processing user data and blockchain work\\
    $\kappa_n$ & The effective switched capacitance that depends on the chip architecture of user~$n$\\
    $\kappa_m$ & The effective switched capacitance that depends on the chip architecture of server~$m$\\
    \bottomrule
\end{tabular}
\vspace{-10pt}
\end{table}

\subsubsection{Trust score function}
The trust score is a quantifiable measure that reflects the reliability or credibility of servers within a blockchain network \cite{al2021blockchain}. This score is calculated by aggregating the evaluated values of one server node over several counts of transactions. These evaluated values are determined based on the outcomes of authentication attempts, both failed and successful, conducted by the respective nodes. The trust score value of the \mbox{$n$-th} user connected to \mbox{$m$-th} server can be denoted as $v_{n,m}$. According to the model introduced in \cite{al2021blockchain}, the total trust score values of all server nodes can be calculated as
\begin{talign}
    \mathcal{V} = \sum\limits_{n \in \mathcal{N}, m \in \mathcal{M}} x_{n,m} v_{n,m}.\label{totaltrust}
\end{talign}\vspace{-2pt}
We further denote the trust score value of user $n$ that is connected to server $m$ as:
\begin{talign}
    &v_{n,m} = \varpi_1 \ln[1+\varpi_2 (\frac{p_{m}}{p_{max}^{(m)}} + \frac{F_{n,m}}{F_{max}^{(m)}} + \frac{b_{n,m}}{b_{max}} + \tau)],\label{trustvalue}
\end{talign}
where $\varpi_1$ and $\varpi_2$ are system-specific constants, and $\tau \in [0,1]$ is a scoring parameter based on the server's history of processing user data and blockchain work. The logarithm function in~(\ref{trustvalue}) above is jointly concave of $p_m$, $F_{n,m}$, and $b_{n,m}$. Using the logarithm function to model the utility has also been adopted by \cite{yang2015incentive} on crowdsensing.
This trust score function is effective and sensitive in all value ranges of $(\frac{p_m}{p_{max}^{(m)}} + \frac{F_{n,m}}{F_{max}^{(m)}} + \frac{b_{n,m}}{b_{max}} + \tau)$, which can describe each user's subjective experience of the communication and computing resources obtained from the server. Some important notation in this paper is shown in Table~\ref{table_notation}.

\subsection{Optimization Problem}

We have computed the system delay, system energy consumption, and the total trust score values in Equations~(\ref{eqTtotal}),~(\ref{eqEtotal}), and~(\ref{totaltrust}) above. Then we define the system trust-cost ratio (TCR) below, which our optimization problem will maximize:
\begin{talign}
    &\frac{\mathcal{V}}{\omega_t T_{\text{total}} + \omega_e E_{\text{total}}}
     = \nonumber \\ &\frac{\sum\limits_{n \in \mathcal{N}, m \in \mathcal{M}} x_{n,m} v_{n,m}}{\omega_t \!\max\limits_{n\in \mathcal{N}, m\in \mathcal{M}}\!\big\{\!\!\max \!\left\{T_{s,n,m}, T_{u,n,m}\right\}\!\big\} + \omega_e (\!\!\sum\limits_{n \in \mathcal{N}} \!\!E_{u,n,m} + \vspace{-10pt}\!\!\!\!\!\sum\limits_{n\in \mathcal{N}, m\in \mathcal{M}}\!\!\!\!\!\!\vspace{-10pt} E_{s,n,m})},\nonumber
\end{talign}
where $\omega_t$ and $\omega_e$ represent the weight values of delay and energy, respectively, with $\omega_t + \omega_e = 1$.

With the above objective function, our optimization variables include the following:
user connection $\bm{x} = [x_{n,m}|_{n \in \mathcal{N}, m \in \mathcal{M} }]$, $\bm{\varphi} = [\varphi_n|_{n \in \mathcal{N}}]$, $\bm{\gamma} = [\gamma_{n,m}|_{n \in \mathcal{N}, m \in \mathcal{M} }]$, bandwidth $\bm{b} = [b_{n,m}|_{n \in \mathcal{N}, m \in \mathcal{M} }]$, transmission power $\bm{p_u} = [p_n|_{n \in \mathcal{N}}]$ and $\bm{p_s} = [p_m|_{m \in \mathcal{M} }]$, and CPU-cycle frequency $\bm{f_u} = [F_n|_{n \in \mathcal{N}}]$ and $\bm{f_s} = [F_{n,m}|_{n \in \mathcal{N}, m \in \mathcal{M} }]$.

\setlength{\abovedisplayskip}{0pt plus 0pt minus 0pt}
\setlength{\belowdisplayskip}{0pt plus 0pt minus 0pt}
\setlength\abovedisplayshortskip{0pt plus 0pt minus 0pt}
\setlength\belowdisplayshortskip{0pt plus 0pt minus 0pt}

Next, we start discussing how to solve the optimization. To begin with, in order to remove the ``maximize" operation in $T_{\text{total}}$, we add an auxiliary variable $T$,  which is constrained to be greater than or equal to both $T_{s,n,m}$ and $T_{u,n,m}$.
Besides, we utilize Dinkelbach's algorithm~\cite{dinkelbach1967nonlinear} to introduce an additional variable $y$, which is obtained from the TCR value in the previous iteration (details later).
Then, the fractional programming in the trust-cost ratio is transformed into the following problem:
\begin{subequations}\label{prob1}
\begin{align}
&\max\limits_{\bm{x},\bm{\varphi},\bm{\gamma},\bm{b},\bm{p_u},\bm{p_s},\bm{f_u},\bm{f_s},T}  \nonumber \\& \hspace{-20pt} \bigg\{ \sum\limits_{n \in \mathcal{N}, m \in \mathcal{M}} [x_{n,m} v_{n,m}  -y \omega_e (E_{u,n,m} + E_{s,n,m})]\bigg\}- y \omega_t T\tag{\ref{prob1}}\\
\text{s.t.} \quad & x_{n,m} \in \{0,1\}, \forall n\in\mathcal{N}, m\in\mathcal{M} \label{x_constr1}\\[-2pt]       & \sum\limits_{m\in \mathcal{M}} x_{n,m} = 1, \forall n\in\mathcal{N} \label{x_constr2}\\
         & \varphi_n \in [0,1], \forall n\in\mathcal{N} \label{varphi_constr}\\
         & \gamma_{n,m} \in (0,1), \forall n\in\mathcal{N}, m\in\mathcal{M} \label{gamma_constr}\\
         & \sum\limits_{n\in \mathcal{N}} x_{n,m} b_{n,m} \leq b_{max},\forall m\in\mathcal{M} \label{x_b_constr}\\
         & p_n \leq p^{(n)}_{max}, \forall n \in\mathcal{N}\label{pu_constr}\\
         & \sum\limits_{n\in \mathcal{N}} x_{n,m} p_{n,m} \leq p^{(m)}_{max}, \forall m\in\mathcal{M} \label{x_ps_constr}\\
         & F_n \leq F^{(n)}_{max}, \forall n\in\mathcal{N} \label{fu_constr}\\
         & \sum\limits_{n\in \mathcal{N}} x_{n,m} F_{n,m} \leq F^{(m)}_{max}, \forall m\in\mathcal{M} \label{x_fs_constr}\\
         & T_{s,n,m} \leq T, \forall n\in\mathcal{N}, m\in\mathcal{M}\label{Ts_constr}\\
         & T_{u,n,m} \leq T, \forall n\in\mathcal{N}, m\in\mathcal{M}\label{Tu_constr}.
\end{align}
\end{subequations}
Based on Dinkelbach's Algorithm, we iteratively optimize $y$ and problem (\ref{prob1}). Specifically, at the $i$-th iteration, given $y^{(i-1)}$, we first obtain $\bm{x}^{(i)},\bm{\varphi}^{(i)},\bm{\gamma}^{(i)},\bm{b}^{(i)},\bm{p_u}^{(i)},\bm{p_s}^{(i)},\bm{f_u}^{(i)},\bm{f_s}^{(i)}, T^{(i)}$ by solving the optimization problem (\ref{prob1}); then we calculate $y^{(i)}$ with the given $\bm{x}^{(i)},\bm{\varphi}^{(i)},\bm{\gamma}^{(i)},\bm{b}^{(i)},\bm{p_u}^{(i)},\bm{p_s}^{(i)},\bm{f_u}^{(i)},\bm{f_s}^{(i)}, T^{(i)}$. Repeat the above operations until the solutions converge. In the following section, we consider using the alternating optimization method (AO) to tackle the complex problem~(\ref{prob1}). 

\textbf{Roadmap of the whole algorithm.} First, we decompose the outer fractional structure of the original TCR problem using the Dinkelbach algorithm and sequentially optimize $\bm{x},\bm{\varphi},\bm{\gamma}$ and $\bm{b},\bm{p_u},\bm{p_s},\bm{f_u},\bm{f_s}$ using the Alternating Optimization (AO) method. In the first step of AO, we fix $\bm{b},\bm{p_u},\bm{p_s},\bm{f_u},\bm{f_s}$ and optimize $\bm{x},\bm{\varphi},\bm{\gamma},T$. We transform the optimization problem in the first step of AO into a Quadratically Constrained Quadratic Program (QCQP) and solve it using Semidefinite Relaxation (SDR) and the Hungarian algorithm. In the second step of AO, we fix $\bm{x},\bm{\varphi},\bm{\gamma}$ and optimize $\bm{b},\bm{p_u},\bm{p_s},\bm{f_u},\bm{f_s}, T$. During the optimization in the second step of AO, we propose a new fractional programming method to transform this non-convex problem into a convex one. Finally, we calculate $y$ based on the obtained solutions and repeat the aforementioned process until $y$ converges. In this algorithm, since we utilize \underline{\textbf{D}}inkelbach's algorithm, \underline{\textbf{a}}lternating optimization, \underline{\textbf{s}}emidefinite relaxation, \underline{\textbf{H}}ungarian algorithm, and \underline{\textbf{f}}ractional programming, we refer to this algorithm as the \textbf{DASHF Algorithm}.

\section{Our proposed AO technique to solve the optimization problem}\label{section.proposed AO technique}
Assuming that $y$ is given, we need to optimize $\bm{x},\bm{\varphi},\bm{\gamma},\bm{b},\bm{p_u},\bm{p_s},\bm{f_u},\bm{f_s}, T$. In the outermost loops, we iteratively optimize $y$; In the innermost loops, we iteratively optimize $\bm{x},\bm{\varphi},\bm{\gamma},\bm{b},\bm{p_u},\bm{p_s},\bm{f_u},\bm{f_s}, T$. However, it is still difficult to optimize them in parallel. Thus, we consider operating two inner AO steps to solve it. At the $i$-th iteration, 
\begin{enumerate}
    \item \textbf{Optimize $\vect{x},\vect{\varphi},\vect{\gamma},T$, given $\vect{b},\vect{p_u},\vect{p_s},\vect{f_u},\vect{f_s}$.} Assuming that $\vect{b}^{(i-1)},\vect{p_u}^{(i-1)},\vect{p_s}^{(i-1)},\vect{f_u}^{(i-1)},\vect{f_s}^{(i-1)},y^{(i-1)}$ are given, we optimize $\vect{x}^{(i)},\vect{\varphi}^{(i)},\vect{\gamma}^{(i)},T^{(i)}$.
    \item \textbf{Optimize $\vect{b},\vect{p_u},\vect{p_s},\vect{f_u},\vect{f_s},T$, given $\vect{x}, \vect{\varphi},\vect{\gamma}$.} Assuming that $\vect{x}^{(i-1)}, \vect{\varphi}^{(i-1)},\vect{\gamma}^{(i-1)},y^{(i-1)}$ are given, we optimize $\vect{b}^{(i)},\vect{p_u}^{(i)},\vect{p_s}^{(i)},\vect{f_u}^{(i)},\vect{f_s}^{(i)},T^{(i)}$.
\end{enumerate}
\subsection{Optimizing $\vect{x}, \vect{\varphi},\vect{\gamma},T$, given $\vect{b},\vect{p_u},\vect{p_s},\vect{f_u},\vect{f_s}$}
Given $\bm{b},\bm{p_u},\bm{p_s},\bm{f_u},\bm{f_s}$, we optimize $\bm{x},\bm{\varphi},\bm{\gamma},T$. The optimization problem will be:
\begin{subequations}\label{prob2}
\begin{align}
\!\!\!\max\limits_{\bm{x},\bm{\varphi},\bm{\gamma},T}\!\!\!\! & \sum\limits_{n \in \mathcal{N}, m \in \mathcal{M}} \!\!\!\!\!\![x_{n,m} v_{n,m} \!-\! y\omega_e (E_{u,n,m} \!+ \!E_{s,n,m})]-y \omega_t T \tag{\ref{prob2}}\\
\text{s.t.}~ & (\text{\ref{x_constr1}}), (\text{\ref{x_constr2}}), (\text{\ref{varphi_constr}}),(\text{\ref{gamma_constr}}),(\text{\ref{x_b_constr}}),(\text{\ref{x_ps_constr}}),(\text{\ref{x_fs_constr}}),(\text{\ref{Ts_constr}}),(\text{\ref{Tu_constr}}).\nonumber
\end{align}
\end{subequations}
Since $x_{n,m}$ are binary and other variables are not discrete, the above belongs to mixed-integer nonlinear programming. We rewrite $x_{n,m}\in \{0,1\}$ as \mbox{$x_{n,m}(x_{n,m}-1)=0$}. The optimization problem will be rewritten as:
\begin{subequations}\label{prob3}
\begin{align}
&\!\!\!\!\max\limits_{\bm{x},\bm{\varphi},\bm{\gamma},T}\!\!\!\!  \sum\limits_{n \in \mathcal{N}, m \in \mathcal{M}} \!\!\!\!\!\!\!\![x_{n,m} v_{n,m} \!-\! y\omega_e (E_{u,n,m} \!+ \!E_{s,n,m})]-y \omega_t T \tag{\ref{prob3}}\\
&\text{s.t.} \quad  x_{n,m}(x_{n,m}-1)=0, \forall n\in\mathcal{N}, m\in\mathcal{M} \label{x_constr1_new}\\
           & \quad\quad(\text{\ref{x_constr2}}), (\text{\ref{varphi_constr}}),(\text{\ref{gamma_constr}}),(\text{\ref{x_b_constr}}),(\text{\ref{x_ps_constr}}),(\text{\ref{x_fs_constr}}),(\text{\ref{Ts_constr}}),(\text{\ref{Tu_constr}}).\nonumber
\end{align}
\end{subequations}
Using the energy expressions of Section~\ref{secEnergy}, we can express the term $E_{u,n,m} + E_{s,n,m}$ in (\ref{prob3}) as 
\begin{talign}
    &\sum\limits_{n \in \mathcal{N}}\kappa_n d_n f_n F^2_n + \sum\limits_{n \in \mathcal{N}} (\omega_p - 1)\varphi_n \kappa_n d_n f_n F^2_n  + \nonumber\\&  \sum\limits_{n \in \mathcal{N}, m \in \mathcal{M}} \{x_{n,m} \varphi_n d_n (\frac{p_n}{r_{n,m}} + \frac{p_m \omega_p}{r_{m,n}}) +\nonumber\\ &  \kappa_m x_{n,m} \varphi_n d_n f_m F^2_{n,m}[\gamma^2_{n,m} + \omega_b (1 - \gamma_{n,m})^2]\},
\end{talign}
where $ \sum_{n \in \mathcal{N}}\kappa_n d_n f_n F^2_n$ is a constant. Since $\bm{b},\bm{p_u},\bm{p_s},\bm{f_u},\bm{f_s}$ are given, $v_{n,m}$ defined in Eq.~(\ref{trustvalue}) is also a constant. Therefore, the objective function in (\ref{prob3}) can be simplified as (\ref{prob4}) below:
\begin{align}
&\!\max\limits_{\bm{x},\bm{\varphi},\bm{\gamma},T}-y\omega_t T +\!\!\!\!\sum\limits_{n \in \mathcal{N}, m \in \mathcal{M}} \!\!\!x_{n,m} v_{n,m} - y \omega_e \bigg\{\!\!\sum\limits_{n \in \mathcal{N}, m \in \mathcal{M}}\!\!\!\! \{x_{n,m} \varphi_n \nonumber \\
&\quad\quad d_n (\frac{p_n}{r_{n,m}} \!+\! \frac{p_m \omega_p}{r_{m,n}})\!+\! \kappa_m x_{n,m} \varphi_n d_n f_m F^2_{n,m}[\gamma^2_{n,m} \!+\! \omega_b (1 \!-\!\nonumber \\ &\quad\quad\gamma_{n,m})^2]\}+ \sum\limits_{n \in \mathcal{N}}\! (\omega_p \!-\! 1)\varphi_n \kappa_n d_n f_n F^2_n\bigg\}\label{prob4}\\
&\text{s.t.}~(\text{\ref{x_constr1_new}}),(\text{\ref{x_constr2}}), (\text{\ref{varphi_constr}}),(\text{\ref{gamma_constr}}),(\text{\ref{x_b_constr}}),(\text{\ref{x_ps_constr}}),(\text{\ref{x_fs_constr}}),(\text{\ref{Ts_constr}}),(\text{\ref{Tu_constr}}).\nonumber
\end{align}
Next, we convert the maximization problem in (\ref{prob4}) to the following minimization problem:
\begin{align}
&\min\limits_{\bm{x},\bm{\varphi},\bm{\gamma},T}\quad y\omega_t T + y \omega_e \bigg\{\sum\limits_{n \in \mathcal{N}} (\omega_p - 1)\varphi_n \kappa_n d_n f_n F^2_n \nonumber \\&\quad\quad\quad\quad+ \hspace{-10pt}\sum\limits_{n \in \mathcal{N}, m \in \mathcal{M}} \{x_{n,m} \varphi_n d_n (\frac{p_n}{r_{n,m}} + \frac{p_m \omega_p}{r_{m,n}})\nonumber \\ &\quad\quad\quad\quad-\hspace{-10pt}\sum\limits_{n \in \mathcal{N}, m \in \mathcal{M}} x_{n,m} v_{n,m} +\kappa_m x_{n,m} \varphi_n d_n f_m F^2_{n,m}[\gamma^2_{n,m} \nonumber \\ &\quad\quad\quad\quad+ \omega_b (1 - \gamma_{n,m})^2]\}\bigg\} \label{prob5}\\
&\text{s.t.} \quad (\text{\ref{x_constr1_new}}),(\text{\ref{x_constr2}}), (\text{\ref{varphi_constr}}),(\text{\ref{gamma_constr}}),(\text{\ref{x_b_constr}}),(\text{\ref{x_ps_constr}}),(\text{\ref{x_fs_constr}}),(\text{\ref{Ts_constr}}),(\text{\ref{Tu_constr}}).\nonumber
\end{align}
The objective function in (\ref{prob5}) is not jointly convex with respect to ($x_{n,m},\varphi_n,\gamma_{n,m}$).
Its hessian determinant: $x_{n,m} \varphi_n (\gamma^2_{n,m} + \omega_b (1 - \gamma_{n,m})^2)(8\gamma_{n,m}-2(\omega_b(1-\gamma_{n,m})^2+\gamma^2_{n,m}))$, which is not always non-negative. Therefore, $x_{n,m} \varphi_n (\gamma^2_{n,m} + \omega_b (1 - \gamma_{n,m})^2)$ is not jointly convex with respect to ($x_{n,m},\varphi_n,\gamma_{n,m}$).

It seems difficult to use the matrix lifting technique to solve this quartic problem by dividing it into several quadratic problems. Then, we seek to find another method to solve it. Let 
\begin{talign}
&A_{n} := y\omega_e (\omega_p - 1)\kappa_n d_n f_n F^2_n,\\
&B_{n,m} \!\!:=  y\omega_e d_n (\frac{p_n}{r_{n,m}} + \frac{p_m \omega_p}{r_{m,n}}) + \omega_b y\omega_e \kappa_m d_n f_m F^2_{n,m},\\
&C_{n,m} := -2\omega_b y\omega_e \kappa_m d_n f_m F^2_{n,m},\\
&D_{n,m} := (\omega_b + 1)y\omega_e \kappa_m d_n f_m F^2_{n,m}.
\end{talign}
Then we define $\mathbf{A} := [A_{n}]|_{n \in \mathcal{N} }$, $\mathbf{B} := [B_{n,m}]|_{n \in \mathcal{N},m\in \mathcal{M} }$, $\mathbf{C} := [C_{n,m}]|_{n \in \mathcal{N},m\in \mathcal{M} }$, and $\mathbf{D} := [D_{n,m}]|_{n \in \mathcal{N},m\in \mathcal{M} }$. We turn to its equivalence problem (\ref{prob6}):
\begin{align}
&\min\limits_{\bm{x},\bm{\varphi},\bm{\gamma},T}y\omega_t T + \sum\limits_{n \in \mathcal{N}} A_{n}\varphi_n + \sum\limits_{n \in \mathcal{N}, m \in \mathcal{M}} (-x_{n,m} v_{n,m}  \nonumber \\ &\quad+ B_{n,m}x_{n,m}\varphi_n + C_{n,m} x_{n,m}\varphi_n \gamma_{n,m} + D_{n,m} x_{n,m}\varphi_n \gamma_{n,m}^2) \label{prob6}\\
&\text{s.t.} ~(\text{\ref{x_constr1_new}}),(\text{\ref{x_constr2}}), (\text{\ref{varphi_constr}}),(\text{\ref{gamma_constr}}),(\text{\ref{x_b_constr}}),(\text{\ref{x_ps_constr}}),(\text{\ref{x_fs_constr}}),(\text{\ref{Ts_constr}}),(\text{\ref{Tu_constr}}).\nonumber
\end{align}

For the polynomial $C_{n,m} x_{n,m}\varphi_n \gamma_{n,m} + D_{n,m} x_{n,m}\varphi_n \gamma_{n,m}^2$ in the objective function of (\ref{prob6}), we can rewrite it as $ x_{n,m}\varphi_n (C_{n,m}\gamma_{n,m} + D_{n,m} \gamma_{n,m}^2)$, where $C_{n,m}<0$ and $D_{n,m}>0$. It means that $C_{n,m}\gamma_{n,m} + D_{n,m} \gamma_{n,m}^2$ is convex, and the minimization point can be obtained when $\gamma_{n,m}=\frac{-C_{n,m}}{2D_{n,m}}=\frac{\omega_b}{\omega_b + 1} \in (0,1)$, which satisfies the constraints of $\gamma_{n,m}$. In $T_{s,n,m}$, the terms $T^{(p_1)}_{n,m}$, $T^{(g)}_{n,m}$, and $T^{(v)}_{n,m}$ are related to $\gamma_{n,m}$. Since $T^{(v)}_{n,m}$ is generally much smaller than $T^{(p_1)}_{n,m}$ and $T^{(g)}_{n,m}$, we only focus on $T^{(p_1)}_{n,m}$ and $T^{(g)}_{n,m}$ here. It's easy to know that when $\gamma_{n,m} = \frac{1}{1+\omega_b}$, $T^{(p_1)}_{n,m} + T^{(g)}_{n,m}$ takes the minimum value according to basic inequality. Following are the detailed steps:
\begin{talign}
    T^{(p_1)}_{n,m} + T^{(g)}_{n,m} \geq 2\sqrt{T^{(p_1)}_{n,m} T^{(g)}_{n,m}},
\end{talign}
where if and only if $T^{(p_1)}_{n,m} =T^{(g)}_{n,m}$, ``$=$'' can be obtained.
\begin{talign}
    &\quad \quad T^{(p_1)}_{n,m} =T^{(g)}_{n,m},\nonumber \\
    &\Rightarrow \frac{x_{n,m}\varphi_n d_n f_m}{\gamma_{n,m}F_{n,m}} = \frac{x_{n,m}\varphi_n d_n\omega_b f_m}{(1-\gamma_{n,m})F_{n,m}}, \nonumber \\
    &\Rightarrow \gamma_{n,m} = \frac{1}{1+\omega_b}.
\end{talign}
We set $\omega_b = 1$ and then $\frac{\omega_b}{1+\omega_b} = \frac{1}{1+\omega_b}$, in which case, the original optimization problem would take the maximum value. Plug in the minimum value of $\gamma_{n,m}$ and the optimization problem can be rewritten as
\begin{subequations}\label{prob7}
\begin{align}
\mathcal{P}_{1}: &\min\limits_{\bm{x},\bm{\varphi},T}\quad  y\omega_t T + \sum\limits_{n \in \mathcal{N}} A_{n}\varphi_n +\!\!\!\!\!\sum\limits_{n \in \mathcal{N}, m \in \mathcal{M}}\!\!\!\!\! -x_{n,m} v_{n,m}  + \nonumber \\ &\quad\quad\quad\quad(\frac{-C_{n,m}^2}{4D_{n,m}}+B_{n,m})x_{n,m}\varphi_n\tag{\ref{prob7}}\\
&\text{s.t.} \quad  (\text{\ref{x_constr1_new}}),(\text{\ref{x_constr2}}), (\text{\ref{varphi_constr}}),(\text{\ref{x_b_constr}}),(\text{\ref{x_ps_constr}}),(\text{\ref{x_fs_constr}}),(\text{\ref{Ts_constr}}),(\text{\ref{Tu_constr}}).\nonumber
\end{align}
\end{subequations}
This is a quadratically constrained quadratic program (QCQP). Then, we need to get the standard form of the QCQP problem. First, we need to combine $\boldsymbol{\varphi}$ and $\boldsymbol{x}$ to define a new vector variable. Therefore, we define the new vector variable as
\begin{talign}
    \boldsymbol{Q}=(\boldsymbol{\varphi}^\intercal,\boldsymbol{x_1}^\intercal,\cdots,\boldsymbol{x_M}^\intercal)^\intercal,
\end{talign}
where $\boldsymbol{\varphi}=(\varphi_1,\cdots,\varphi_N)^\intercal$ and $\boldsymbol{x_m}=(x_{1,m},\cdots,x_{N,m})^\intercal, \forall m\in\mathcal{M}$. To make the problem transformation more clearly, we define some auxiliary vectors and matrices as follows:
\begin{talign}
    &\boldsymbol{e}_i = (0,\cdots,\underset{\underset{i\text{-th}}{\uparrow}}{1},\cdots,0)^\intercal_{NM+N \times 1},\\
    &\boldsymbol{e}_{i,j} = (\boldsymbol{e}_i,\cdots,\boldsymbol{e}_j)^\intercal, i<j, \\
    &\boldsymbol{I}_{NM+N\times N}=(\boldsymbol{I}_N, \boldsymbol{0}_{N\times NM+N})^\intercal, \\
    &\boldsymbol{I}_{N\rightarrow NM}=(\boldsymbol{I}_N,\cdots,\boldsymbol{I}_N)_{N\times NM},\\
    &\boldsymbol{e}_{N+1\times N+NM}=(\boldsymbol{0}_{NM\times N},\boldsymbol{I}_{NM}),\\
     &\boldsymbol{e}_{\overline{i}}=(0,\cdots,\underset{\underset{i\text{-th}}{\uparrow}}{1},\cdots,\underset{\underset{(i+N)\text{-th}}{\uparrow}}{1},\cdots,\underset{\underset{(i+N(M-1))\text{-th}}{\uparrow}}{1},\cdots,0)^\intercal,
\end{talign}
where $\boldsymbol{I}_N$ is the identity matrix of order $N$ and $\boldsymbol{I}_{NM}$ is the identity matrix of order $NM$. We also define
\begin{talign}
    &G_{n,m}=\frac{-C_{n,m}^2}{4D_{n,m}}+B_{n,m},\\
    &\boldsymbol{G}=(G_{1,1},\cdots,G_{N,M})^\intercal.
\end{talign}
Thus, we can obtain that
\begin{talign}
    &\sum_{n\in \mathcal{N},m\in \mathcal{M}}(\frac{-C_{n,m}^2}{4D_{n,m}}+B_{n,m})x_{n,m}\varphi_n \nonumber\\
    &=\boldsymbol{Q}^\intercal \boldsymbol{I}_{NM+N\times N}I_{N\rightarrow NM}\text{diag}(\boldsymbol{G})\boldsymbol{e}_{N+1\times N+NM}\boldsymbol{Q}.
\end{talign}
Let 
\begin{equation}
\mathbf{P}_0=\boldsymbol{I}_{NM+N\times N}I_{N\rightarrow NM}\text{diag}(\boldsymbol{G})\boldsymbol{e}_{N+1\times N+NM}.
\end{equation}
Similarly, we also know that
\begin{talign}
    \sum_{n \in \mathcal{N}, m \in \mathcal{M}} -x_{n,m} v_{n,m}=\mathbf{W}_0^\intercal\boldsymbol{Q},
\end{talign}
where 
\begin{talign}
    \mathbf{W}_0^\intercal=(-\boldsymbol{v}_1^\intercal,\cdots,-\boldsymbol{v}_M^\intercal)\boldsymbol{e}_{N+1,NM+N},
\end{talign}
and $\boldsymbol{v}_i^\intercal=(v_{1,i},\cdots,v_{N,i}), \forall i \in \mathcal{M}$. In addition, $\sum_{n \in \mathcal{N}, m \in \mathcal{M}} A_{n}\varphi_n$ can be expressed as $\mathbf{W}_1^\intercal\boldsymbol{Q}$, where $\mathbf{W}_1^\intercal=(A_1,\cdots,A_N)\boldsymbol{e}_{1,N}$. Let
\begin{talign}
    &P^{(\text{Ts})}_{1,n,m} = \frac{d_n}{r_{n,m}} + \frac{d_n f_m}{\gamma_{n,m} F_{n,m}} + \frac{d_n \omega_b f_m}{(1-\gamma_{n,m})F_{n,m}},\\
    &P^{(\text{Ts})}_2 = \frac{S_b}{R_{m}} + \mathop{\max}_{m{'} \in \mathcal{M} \setminus \{m\}} \frac{f_v}{(1-\gamma_{n,m^\prime})F_{n,m^\prime}},\\
    &P^{(\text{Tu})}_{1,n,m} = \frac{d_n \omega_p}{r_{m,n}},\\
    &P^{(\text{Tu})}_{2,n} = \frac{d_n \omega_p f_n - d_n f_n}{F_n},\\
    &P^{(\text{Tu})}_{3} = \frac{d_n f_n}{F_n}.
\end{talign}
Furthermore, let $\mathbf{P}^{(\text{Ts})}_{3} = [P^{(\text{Ts})}_{1,n,m}]|_{n \in \mathcal{N},m\in \mathcal{M} }$, 
\begin{talign}
    \mathbf{P}^{(\text{Ts})}_1=\boldsymbol{I}_{NM+N\times N}I_{N\rightarrow NM}\text{diag}(\mathbf{P}^{(\text{Ts})}_{3})\boldsymbol{e}_{N+1\times N+NM},
\end{talign}
$\mathbf{P}^{(\text{Tu})}_4 = [P^{(\text{Tu})}_{1,n,m}]|_{n \in \mathcal{N},m\in \mathcal{M} }$, and 
\begin{talign}
\mathbf{P}^{(\text{Tu})}_1=\boldsymbol{I}_{NM+N\times N}I_{N\rightarrow NM}\text{diag}(\mathbf{P}^{(\text{Tu})}_4)\boldsymbol{e}_{N+1\times N+NM}.
\end{talign}
Let $\mathbf{P}^{(\text{Tu})}_5 = [P^{(\text{Tu})}_{2,n}]|_{n \in \mathcal{N}}$ and ${\mathbf{P}^{(\text{Tu})}_2}^\intercal = {\mathbf{P}^{(\text{Tu})}_5}^\intercal\boldsymbol{e}_{1,N}$. 
Based on the above, the optimization problem can be expressed as
\begin{subequations}\label{prob8}
\begin{align}
\mathcal{P}_{2}: &\min\limits_{\bm{Q},T}\quad  \boldsymbol{Q}^\intercal\mathbf{P}_0 \boldsymbol{Q}+\mathbf{W}_{0}^\intercal\boldsymbol{Q}+\mathbf{W}_{1}^\intercal\boldsymbol{Q}+y\omega_t T\tag{\ref{prob8}}\\
\text{s.t.} \quad & \text{diag}(\boldsymbol{e}_{N+1,NM+N}^\intercal\boldsymbol{Q})(\text{diag}(\boldsymbol{e}_{N+1,NM+N}^\intercal\boldsymbol{Q})-\mathbf{I})=\mathbf{0} \label{x_constr1_qcqp}\\       & \text{diag}(\boldsymbol{e}_{\overline{1},\overline{M}}^\intercal\boldsymbol{e}_{N+1,NM+N}^\intercal\boldsymbol{Q})=\mathbf{I} \label{x_constr2_qcqp}\\
         & \text{diag}(\boldsymbol{e}_{1,N}^\intercal\boldsymbol{Q})\leq\mathbf{I} \label{varphi_constr1_qcqp}\\
         & \text{diag}(\boldsymbol{e}_{1,N}^\intercal\boldsymbol{Q})\geq\mathbf{0} \label{varphi_constr2_qcqp}\\
         & \boldsymbol{B}^\intercal\boldsymbol{e}_{N+1,NM+N}^\intercal\boldsymbol{Q}-B_{max} \leq 0 \label{x_b_constr_qcqp}\\
         & \boldsymbol{P}^\intercal\boldsymbol{e}_{N+1,NM+N}^\intercal\boldsymbol{Q}-P_{max}^{(m)} \leq 0 \label{x_ps_constr_qcqp}\\
         & \boldsymbol{F}^\intercal\boldsymbol{e}_{N+1,NM+N}^\intercal\boldsymbol{Q}-F_{max}^{(m)} \leq 0 \label{x_fs_constr_qcqp}\\
         & \boldsymbol{Q}^\intercal\mathbf{P}^{(\text{Ts})}_1 \boldsymbol{Q} + P^{(\text{Ts})}_2 \preceq T \label{Ts_constr_qcqp}\\
         & \boldsymbol{Q}^\intercal\mathbf{P}^{(\text{Tu})}_1 \boldsymbol{Q} + {\mathbf{P}^{(\text{Tu})}_{2}}^\intercal \boldsymbol{Q} + P^{(\text{Tu})}_3 \preceq T, \label{Tu_constr_qcqp}
\end{align}
\end{subequations}
where $\boldsymbol{e}_{\overline{i},\overline{j}}$ denotes $(\boldsymbol{e}_{\overline{i}},\cdots,\boldsymbol{e}_{\overline{j}})^\intercal, i<j$, $\boldsymbol{B}=(b_{1,1},\cdots,b_{N,M})^\intercal$, $\boldsymbol{P}=(p_{1,1},\cdots,p_{N,M})^\intercal$, and $\boldsymbol{F}=(F_{1,1},\cdots,F_{N,M})^\intercal$. 
The constraints $(\text{\ref{x_constr1_new}})$, $(\text{\ref{x_constr2}})$, $(\text{\ref{varphi_constr}})$, $(\text{\ref{x_b_constr}})$, $(\text{\ref{x_ps_constr}})$, $(\text{\ref{x_fs_constr}})$, $(\text{\ref{Ts_constr}})$, $(\text{\ref{Tu_constr}})$ in Problem $\mathcal{P}_{1}$ are transformed into the constraints $(\text{\ref{x_constr1_qcqp}})$, $(\text{\ref{x_constr2_qcqp}})$, $\{(\text{\ref{varphi_constr1_qcqp}}), (\text{\ref{varphi_constr2_qcqp}})\}$,  $(\text{\ref{x_b_constr_qcqp}})$, $(\text{\ref{x_ps_constr_qcqp}})$, $(\text{\ref{x_fs_constr_qcqp}})$, $(\text{\ref{Ts_constr_qcqp}})$, $(\text{\ref{Tu_constr_qcqp}})$ in Problem $\mathcal{P}_{2}$, respectively. Problem (\ref{prob8}) is the standard QCQP form. However, it is still non-convex. Then, we need to utilize the semidefinite programming (SDP) method to transform this QCQP problem into a semidefinite relaxation (SDR) problem. Let $\mathbf{S}=(\vect{Q}^\intercal,1)^\intercal(\vect{Q}^\intercal,1)$. Let $\boldsymbol{e}_{i \rightarrow j}$ denotes $(0,\cdots,\underset{\underset{i\text{-th}}{\uparrow}}{1},1,\cdots,\underset{\underset{j\text{-th}}{\uparrow}}{1},0,\cdots,0)^\intercal, i<j$. Then we obtain the SDR problem
\begin{subequations}\label{prob9}
\begin{align}
\mathcal{P}_{3}: \min\limits_{\bm{S},T}\quad & \text{Tr}(\mathbf{P}_1 \mathbf{S})\tag{\ref{prob9}}\\
\text{s.t.} \quad & \text{Tr}(\mathbf{P}_2 \mathbf{S})=0 \label{x_constr1_sdr}\\       & \text{Tr}(\mathbf{P}_3 \mathbf{S})=0 \label{x_constr2_sdr}\\
         & \text{Tr}(\mathbf{P}_4 \mathbf{S})\leq0 \label{varphi_constr_sdr}\\
         & \text{Tr}(\mathbf{P}_5 \mathbf{S})\leq0 \label{x_b_constr_sdr}\\
         & \text{Tr}(\mathbf{P}_6 \mathbf{S})\leq0 \label{x_ps_constr_sdr}\\
         & \text{Tr}(\mathbf{P}_7 \mathbf{S})\leq0 \label{x_fs_constr_sdr}\\
         & \text{Tr}(\mathbf{P}_8 \mathbf{S})\preceq T \label{Ts_constr_sdr}\\
         & \text{Tr}(\mathbf{P}_9 \mathbf{S})\preceq T \label{Tu_constr_sdr}\\
         & \mathbf{S}\succeq0, \label{S_constr_sdr}
\end{align}
\end{subequations}
where 
\begin{equation}
\mathbf{P}_1=
\left(
    \begin{array}{cc}
       \mathbf{P}_0  & \frac{1}{2}(\mathbf{W}_0+\mathbf{W}_1) \\
        \frac{1}{2}(\mathbf{W}_0+\mathbf{W}_1)^\intercal & y\omega_t T
    \end{array}
\right), \nonumber
\end{equation}
\begin{equation}
\mathbf{P}_2=
\left(
    \begin{array}{cc}
      \boldsymbol{e}_{i}^\intercal\boldsymbol{e}_{i}   & -\frac{1}{2}\boldsymbol{e}_{i} \\
       -\frac{1}{2}\boldsymbol{e}_{i}^\intercal  & 0
    \end{array}
\right), \forall i \in \{1,\cdots, NM\}\nonumber
\end{equation}
\begin{align}
\mathbf{P}_3=
\left(
    \begin{array}{cc}
    \mathbf{0}_{NM+N \times NM+N}     & \frac{1}{2}(\boldsymbol{e}_{\overline{i}}\boldsymbol{e}_{N+1,NM+N}^\intercal) \\
     \frac{1}{2}(\boldsymbol{e}_{\overline{i}}\boldsymbol{e}_{N+1,NM+N}^\intercal)^\intercal    & -1
    \end{array}
\right)\nonumber, \\ \forall i \in \{1,\cdots, N\}\nonumber
\end{align}
\begin{equation}
\mathbf{P}_4=
\left(
    \begin{array}{cc}
      \mathbf{0}_{NM+N \times NM+N}   & \frac{1}{2}\boldsymbol{e}_{i} \\
       \frac{1}{2}\boldsymbol{e}_{i}^\intercal  & -1
    \end{array}
\right), \forall i \in \{1,\cdots, N\}\nonumber
\end{equation}
\begin{equation}
\mathbf{P}_5=
\left(
    \begin{array}{cc}
    \mathbf{0}_{NM+N \times NM+N}    & \frac{1}{2}\boldsymbol{B}\boldsymbol{e}_{N+1,NM+N} \\
    \frac{1}{2}(\boldsymbol{B}\boldsymbol{e}_{N+1,NM+N})^\intercal     & -B_{max}
    \end{array}
\right),\nonumber
\end{equation}
\begin{equation}
\mathbf{P}_6=
\left(
    \begin{array}{cc}
    \mathbf{0}_{NM+N \times NM+N}     & \frac{1}{2}\boldsymbol{P}\boldsymbol{e}_{N+1,NM+N} \\
    \frac{1}{2}(\boldsymbol{P}\boldsymbol{e}_{N+1,NM+N})^\intercal     & -P_{max}^{(m)}
    \end{array}
\right),\nonumber
\end{equation}
\begin{equation}
\mathbf{P}_7=
\left(
    \begin{array}{cc}
    \mathbf{0}_{NM+N \times NM+N}     & \frac{1}{2}\boldsymbol{F}\boldsymbol{e}_{N+1,NM+N} \\
    \frac{1}{2}(\boldsymbol{F}\boldsymbol{e}_{N+1,NM+N})^\intercal     & -F_{max}^{(m)}
    \end{array}
\right), \nonumber
\end{equation}
\begin{equation}
\mathbf{P}_8=
\left(
    \begin{array}{cc}
    \mathbf{P}_1^{(\text{Ts})}     & \mathbf{0}_{NM+N \times 1}\\
    \mathbf{0}_{1 \times NM+N}     & P_2^{(\text{Ts})}
    \end{array}
\right), \nonumber
\end{equation}
\begin{equation}
\mathbf{P}_9=
\left(
    \begin{array}{cc}
    \mathbf{P}_1^{(\text{Tu})}     & \frac{1}{2}\vect{P}_2^{(\text{Tu})} \\
    \frac{1}{2}{\vect{P}_2^{(\text{Tu})}}^\intercal     & P_3^{(\text{Tu})}
    \end{array}
\right). \nonumber
\end{equation}
\begin{algorithm*}[h]
\caption{AO-Part1: Optimizing $\vect{x}, \vect{\varphi}, T$, Given $\vect{b},\vect{p_u},\vect{p_s},\vect{f_u},\vect{f_s}$.}
\label{algo:AO-part1}

Initialize $i \leftarrow -1$ and for all $n \in \mathcal{N}, m \in \mathcal{M}$: $\vect{x}^{(0)} = (\vect{e_1},\cdots,\vect{e_M})^\intercal$, $\varphi_n^{(0)}=0.5$, $\gamma_{n,m} = \frac{\omega_b}{\omega_b + 1}$;

Calculate $y^{(0)}, T^{(0)},\mathbf{A}^{(0)},\mathbf{B}^{(0)},\mathbf{C}^{(0)},\mathbf{D}^{(0)},\mathbf{G}^{(0)},\mathbf{P}^{(0)}_1,\mathbf{P}^{(0)}_2,\mathbf{P}^{(0)}_3,\mathbf{P}^{(0)}_4,\mathbf{P}^{(0)}_5,\mathbf{P}^{(0)}_6,\mathbf{P}^{(0)}_7,\mathbf{P}^{(0)}_8,\mathbf{P}^{(0)}_9$ with $\vect{x}^{(0)} = (\vect{e_1},\cdots,\vect{e_M})^\intercal$, $\varphi_n^{(0)}=0.5$, $\gamma_{n,m} = \frac{\omega_b}{\omega_b + 1}$;

\Repeat{the relative difference between $V_{\text{AO-P1}}(\vect{x}^{(i+1)}, \vect{\varphi}^{(i+1)})$ and $V_{\text{AO-P1}}(\vect{x}^{(i)}, \vect{\varphi}^{(i)})$ is no greater than $\epsilon_3$ for a small positive number $\epsilon_3$ (i.e., $\frac{V_{\text{AO-P1}}(\vect{x}^{(i+1)}, \vect{\varphi}^{(i+1)})}{V_{\text{AO-P1}}(\vect{x}^{(i)}, \vect{\varphi}^{(i)})}- 1 \leq \epsilon_3$)}{

Let $i \leftarrow i+1$;

Initialize $j = -1$, $[\vect{x}^{(i,0)},\vect{\varphi}^{(i,0)}] \leftarrow [\vect{x}^{(i)},\vect{\varphi}^{(i)}]$;

Initialize $ [y^{(i,0)},T^{(i,0)}, \mathbf{A}^{(i,0)},\mathbf{B}^{(i,0)},\mathbf{C}^{(i,0)},\mathbf{D}^{(i,0)},\mathbf{G}^{(i,0)},\mathbf{P}^{(i,0)}_1,\mathbf{P}^{(i,0)}_2,\mathbf{P}^{(i,0)}_3,\mathbf{P}^{(i,0)}_4,\mathbf{P}^{(i,0)}_5,\mathbf{P}^{(i,0)}_6,\mathbf{P}^{(i,0)}_7,\mathbf{P}^{(i,0)}_8,\mathbf{P}^{(i,0)}_9] \leftarrow [y^{(i)},T^{(i)} \mathbf{A}^{(i)},\mathbf{B}^{(i)},\mathbf{C}^{(i)},\mathbf{D}^{(i)},\mathbf{G}^{(i)},\mathbf{P}^{(i)}_1,\mathbf{P}^{(i)}_2,\mathbf{P}^{(i)}_3,\mathbf{P}^{(i)}_4,\mathbf{P}^{(i)}_5,\mathbf{P}^{(i)}_6,\mathbf{P}^{(i)}_7,\mathbf{P}^{(i)}_8,\mathbf{P}^{(i)}_9]$

\Repeat{the relative difference between $V_{\text{SDR}}(\vect{x}^{(i,j+1)}, \vect{\varphi}^{(i,j+1)})$ and $V_{\text{SDR}}(\vect{x}^{(i,j)}, \vect{\varphi}^{(i,j)})$ is no greater than $\epsilon_1$ for a small positive number $\epsilon_1$ (i.e., $\frac{V_{\text{SDR}}(\vect{x}^{(i,j+1)}, \vect{\varphi}^{(i,j+1)})}{V_{\text{SDR}}(\vect{x}^{(i,j)}, \vect{\varphi}^{(i,j)})}- 1 \leq \epsilon_1$)}{
Let $j \leftarrow j+1$;

Obtain $[\vect{x}^{(i,j+1)}, \vect{\varphi}^{(i,j+1)}]$ of continuous values by solving the SDR Problem (\ref{prob9});

Update $[y^{(i,j+1)}, T^{(i,j+1)},\mathbf{A}^{(i,j+1)},\mathbf{B}^{(i,j+1)},\mathbf{C}^{(i,j+1)},\mathbf{D}^{(i,j+1)},\mathbf{G}^{(i,j+1)},\mathbf{P}^{(i,j+1)}_1,\mathbf{P}^{(i,j+1)}_2,\mathbf{P}^{(i,j+1)}_3,\mathbf{P}^{(i,j+1)}_4,\mathbf{P}^{(i,j+1)}_5,$ $\mathbf{P}^{(i,j+1)}_6,\mathbf{P}^{(i,j+1)}_7,\mathbf{P}^{(i,j+1)}_8,\mathbf{P}^{(i,j+1)}_9]$;

}
Denote $\vect{x}^{(i,j+1)},\vect{\varphi}^{(i,j+1)}$ as a solution to the SDR Problem (\ref{prob9});

Find all user $n$ that $\sum\limits_{m \in \mathcal{M}} x_{n,m} > 1$. For these users, modify $x_{n,m}$ as $\frac{x_{n,m}}{|\sum\limits_{m \in \mathcal{M}} x_{n,m}|}$. Use the Hungarian algorithm with augmented zero vectors to find the best matching with the maximum edge weight. Denote this matching as a set $\mathcal{X}_{matching}$. For nodes $n$ and $m$ in $\mathcal{X}_{matching}$, let $x_{n,m} = 1$, else $x_{n,m} = 0$, and denote this integer association result as $\vect{x}^{(i,j+1)}_{\sharp}$.

Initialize $l = -1$, $[\vect{x}^{(i,j+1,0)},\vect{\varphi}^{(i,j+1,0)}] \leftarrow [\vect{x}^{(i,j+1)}_{\sharp},\vect{\varphi}^{(i,j+1)}]$;

Initialize $[y^{(i,j+1,0)}, T^{(i,j+1,0)},\mathbf{A}^{(i,j+1,0)},\mathbf{B}^{(i,j+1,0)},\mathbf{C}^{(i,j+1,0)},\mathbf{D}^{(i,j+1,0)},\mathbf{G}^{(i,j+1,0)}] \leftarrow [y^{(i,j+1)}, \mathbf{A}^{(i,j+1)},\mathbf{B}^{(i,j+1)},\mathbf{C}^{(i,j+1)},\mathbf{D}^{(i,j+1)},\mathbf{G}^{(i,j+1)}]$;

\Repeat{the relative difference between $V_{\text{Eq(\ref{prob7})}}(\vect{\varphi}^{(i,j+1,l+1)})$ and $V_{\text{Eq(\ref{prob7})}}(\vect{\varphi}^{(i,j+1,l)})$ is no greater than $\epsilon_2$ for a small positive number $\epsilon_2$ (i.e., $\frac{V_{\text{Eq(\ref{prob7})}}(\vect{\varphi}^{(i,j+1,l+1)})}{V_{\text{Eq(\ref{prob7})}}(\vect{\varphi}^{(i,j+1,l)})}- 1 \leq \epsilon_2$)}{
Let $l \leftarrow l+1$;

Obtain $\vect{\varphi}^{(i,j+1,l+1)}$ by solving Problem (\ref{prob7}) with $\vect{x}^{(i,j+1,0)}$, and denote $\vect{\varphi}^{(i,j+1,l+1)}$ as a solution to Problem (\ref{prob9});

Update $[y^{(i,j+1,l+1)}, T^{(i,j+1,l+1)},\mathbf{A}^{(i,j+1,l+1)},\mathbf{B}^{(i,j+1,l+1)},\mathbf{C}^{(i,j+1,l+1)},\mathbf{D}^{(i,j+1,l+1)},\mathbf{G}^{(i,j+1,l+1)}]$;
}

Set $[\vect{x}^{i+1},\vect{\varphi}^{i+1}] \leftarrow [\vect{x}^{(i,j+1,0)},\vect{\varphi}^{(i,j+1,l+1)}]$;
}

Calculate $T^{(i+1)}$ based on $[\boldsymbol{x}^{(i+1)},\boldsymbol{\varphi}^{(i+1)}]$;

Return $[\boldsymbol{x}^{(i+1)},\boldsymbol{\varphi}^{(i+1)},T^{(i+1)}]$ as a solution $[\boldsymbol{x}^{\star},\boldsymbol{\varphi}^{\star},T^{\star}]$ to Problem $\text{AO-Part1}$.
\end{algorithm*}
The constraints $(\text{\ref{x_constr1_qcqp}})$, $(\text{\ref{x_constr2_qcqp}})$, $(\text{\ref{varphi_constr1_qcqp}})$, $(\text{\ref{varphi_constr2_qcqp}})$, $(\text{\ref{x_b_constr_qcqp}})$, $(\text{\ref{x_ps_constr_qcqp}})$, $(\text{\ref{x_fs_constr_qcqp}})$, $(\text{\ref{Ts_constr_qcqp}})$, $(\text{\ref{Tu_constr_qcqp}})$ in $\mathcal{P}_{2}$ are transformed into the constraints $(\text{\ref{x_constr1_sdr}})$, $(\text{\ref{x_constr2_sdr}})$, $(\text{\ref{varphi_constr_sdr}})$, $(\text{\ref{S_constr_sdr}})$, $(\text{\ref{x_b_constr_sdr}})$, $(\text{\ref{x_ps_constr_sdr}})$, $(\text{\ref{x_fs_constr_sdr}})$, $(\text{\ref{Ts_constr_sdr}})$, $(\text{\ref{Tu_constr_sdr}})$ in $\mathcal{P}_{3}$, respectively. We drop the constraint $\text{rank}(\mathbf{S})=1$, and the objective function and the constraints are all convex. Then, this SDR problem will be solved in polynomial time by common convex solvers. By solving this SDR problem, we can get a continuous solution of $\vect{Q}$. However, this solution is the lower bound of the optimal solution, and it may not guarantee the constraint $\text{rank}(\mathbf{S})=1$. Therefore, we need to use rounding techniques to recover the solution. The latter $NM$ elements in $\vect{Q}$ are $x_{n,m}$, for all $n \in \mathcal{N}, m \in \mathcal{M}$, which means that user $n$ is fractional connected to server $m$. Then, we find all user $n$ that $\sum\limits_{m \in \mathcal{M}} x_{n,m} > 1$. For these users, we modify $x_{n,m}$ as $\frac{x_{n,m}}{|\sum\limits_{m \in \mathcal{M}} x_{n,m}|}$. By using the Hungarian algorithm \cite{dai2018joint} with augmented zero vectors, we find the best matching with the maximum weight and denote this matching as a set $\mathcal{X}_{matching}$. For nodes $n$ and $m$ in $\mathcal{X}_{matching}$, let $x_{n,m} = 1$, else $x_{n,m} = 0$, and denote this integer association result as $\vect{x}_{\sharp}$. Then, substitute $\vect{x}_{\sharp}$ into Problem (\ref{prob8}) to obtain the optimal $\vect{\varphi}$. The process of Algorithm AO-Part 1 is presented in \textbf{Algorithm~\ref{algo:AO-part1}}.

\subsection{Optimizing $\vect{b},\vect{p_u},\vect{p_s},\vect{f_u},\vect{f_s},T$, given $\vect{x}, \vect{\varphi},\vect{\gamma}$}
Given $\vect{x}, \vect{\varphi},\vect{\gamma}$, the remaining optimization problem is
\begin{align}
&\max\limits_{\bm{b},\bm{p_u},\bm{p_s},\bm{f_u},\bm{f_s},T}\quad \sum\limits_{n \in \mathcal{N}, m \in \mathcal{M}} x_{n,m} v_{n,m} - y \bigg\{\omega_t T \nonumber \\&+ \omega_e \big\{\sum\limits_{n \in \mathcal{N}} [\kappa_n (1-\varphi_n) d_n f_n F^{2}_n + \kappa_n \varphi_n d_n \omega_p f_n F_n^2] \nonumber\\& +\sum\limits_{n\in \mathcal{N}, m\in \mathcal{M}} \{p_n \frac{x_{n,m}\varphi_n d_n}{r_{n,m}} +  \kappa_m x_{n,m} \varphi_n d_n f_m (\gamma_{n,m} F_{n,m})^{2} \nonumber \\&+ \hspace{-2pt}\kappa_m x_{n,m} \varphi_n d_n \omega_b f_m [(1\hspace{-2pt}-\hspace{-2pt}\gamma_{n,m})F_{n,m}]^2 \hspace{-2pt}+\hspace{-2pt}  p_m \frac{x_{n,m} \varphi_n d_n \omega_p}{r_{m,n}}\hspace{-2pt}\}\hspace{-2pt}\big\}\hspace{-2pt}\bigg\}\label{prob10}\\
&\text{s.t.} \quad 
         (\text{\ref{x_b_constr}}), (\text{\ref{pu_constr}}), (\text{\ref{x_ps_constr}}), (\text{\ref{fu_constr}}), (\text{\ref{x_fs_constr}}), (\text{\ref{Ts_constr}}), (\text{\ref{Tu_constr}}).\nonumber
\end{align}

\setlength{\abovedisplayskip}{0pt plus 0pt minus 0pt}
\setlength{\belowdisplayskip}{0pt plus 0pt minus 0pt}
\setlength\abovedisplayshortskip{0pt plus 0pt minus 0pt}
\setlength\belowdisplayshortskip{0pt plus 0pt minus 0pt}

We define
\begin{talign}
    &\mathcal{F}(\bm{b},\bm{p_s},\bm{f_u},\bm{f_s},T):=\!\!\!\!\sum\limits_{n \in \mathcal{N}, m \in \mathcal{M}}\!\!x_{n,m} v_{n,m}  \nonumber \\& -y\omega_e \big\{\sum\limits_{n \in \mathcal{N}} [\kappa_n (1-\varphi_n) d_n f_n F^{2}_n + \kappa_n \varphi_n d_n \omega_p f_n F_n^2 ] \nonumber\\& + \sum\limits_{n\in \mathcal{N}, m\in \mathcal{M}} \!\!\!\!\!\{\kappa_m x_{n,m} \varphi_n d_n f_m (\gamma_{n,m} F_{n,m})^{2}  \nonumber \\& + \kappa_m x_{n,m} \varphi_n d_n \omega_b f_m [(1-\gamma_{n,m})F_{n,m}]^2 \}\big\}-y \omega_t T.
\end{talign}
It's easy to justify that $\mathcal{F}(\bm{b},\bm{p_s},\bm{f_u},\bm{f_s},T)$ is concave. Then, the optimization objective function is 
\begin{talign}\label{fp_ori_prob}
    \mathcal{F}(\bm{b},\bm{p_s},\bm{f_u},\bm{f_s},T)\!-\!y \omega_e \!\!\!\!\!\!\sum\limits_{n \in \mathcal{N}, m \in \mathcal{M}}\!\!\!\!\!\!\!(\frac{p_n x_{n,m}\varphi_n d_n}{r_{n,m}} \!+\! \frac{p_m x_{n,m} \varphi_n d_n \omega_p}{r_{m,n}}),
\end{talign}
where $\frac{p_n x_{n,m}\varphi_n d_n}{r_{n,m}} + \frac{p_m x_{n,m} \varphi_n d_n \omega_p}{r_{m,n}}$ is non-convex or concave. Then, according to the fractional programming technique introduced in Section \uppercase\expandafter{\romannumeral4} in \cite{zhao2023human}, we let $z_{1,n,m}=\frac{1}{2 p_n x_{n,m}\varphi_n d_n r_{n,m}}$ and $z_{2,n,m}=\frac{1}{2 p_m x_{n,m} \varphi_n d_n \omega_p r_{m,n}}$. The optimization objective function can be expressed as
\begin{talign} 
    &\mathcal{F}(\bm{b},\bm{p_s},\bm{f_u},\bm{f_s},T)-y \omega_e \sum\limits_{n \in \mathcal{N}, m \in \mathcal{M}}[(p_n x_{n,m}\varphi_n d_n)^2z_{1,n,m}\nonumber \\&+\frac{1}{4 (b_{n,m}\log_2(1+\frac{g_{n,m}p_{n}}{\sigma^2b_{n,m}}))^2 z_{1,n,m}} + (p_m x_{n,m} \varphi_n d_n \omega_p)^2 z_{2,n,m} \nonumber\\ &+ \frac{1}{4 (b_{m,n}\log_2(1+\frac{g_{m,n}p_{m}}{\sigma^2b_{m,n}}))^2 z_{2,n,m}}] \label{fp_trans_prob}
\end{talign}
The transformed optimization problem is shown as (\ref{prob11}):
\begin{align}
&\max\limits_{\bm{b},\bm{p_u},\bm{p_s},\bm{f_u},\bm{f_s},T} \text{The expression in (\ref{fp_trans_prob}) above}\label{prob11}\\
& \quad \text{s.t. }
         (\text{\ref{x_b_constr}}), (\text{\ref{pu_constr}}), (\text{\ref{x_ps_constr}}), (\text{\ref{fu_constr}}), (\text{\ref{x_fs_constr}}),  (\text{\ref{Ts_constr}}), (\text{\ref{Tu_constr}}).\nonumber
\end{align}
If $\vect{z}_1 = (z_{1,1,1},\cdots,z_{1,n,m})^\intercal$ and $\vect{z}_2 = (z_{2,1,1},\cdots,z_{2,n,m})^\intercal$ is given, the objective function (\ref{prob11}) is concave. At the $i$-th iteration, $[y^{(i)}$, $\vect{z}_1^{(i)}$, $\vect{z}_2^{(i)}]$ are first calculated with the solution $[\vect{b}^{(i-1)},\vect{p_u}^{(i-1)},\vect{p_s}^{(i-1)},\vect{f_u}^{(i-1)},\vect{f_s}^{(i-1)}]$. Then, $[\vect{b}^{(i)},\vect{p_u}^{(i)},\vect{p_s}^{(i)},\vect{f_u}^{(i)},\vect{f_s}^{(i)}]$ can be obtained by solving the concave problem (\ref{prob11}) with $[y^{(i)}, \vect{z}_1^{(i)}, \vect{z}_2^{(i)}]$. Thus, the optimization problem is concave and can be solved by common convex solvers. Denote the value of the objective function is $V_{\text{AO-Part2}}$. The process of FP is presented in \textbf{Algorithm~\ref{algo:AO-part2}}. The whole process of the DASHF algorithm is presented in \textbf{Algorithm~\ref{algo:AO-p3}}.

\subsubsection{Prove that the maximization optimization of (\ref{prob10}) is equal to the maximization optimization of (\ref{prob11})}
Let $\mathit{M}_1 (p_n) = p_n x_{n,m}\varphi_n d_n$, $\mathit{M}_2 (p_m) = p_m x_{n,m}\varphi_n d_n \omega_p$, $\mathit{N}_1 (b_{n,m}, p_n) = r_{n,m}$, and $\mathit{N}_2 (b_{n,m}, p_m) = r_{m,n}$. We know that $\mathit{M}_1 (p_n)$ and $\mathit{M}_2 (p_m)$ are convex of $p_n$ and $p_m$, $\mathit{N}_1 (b_{n,m}, p_n)$ and $\mathit{N}_2 (b_{n,m}, p_m)$ are jointly concave of $(b_{n,m}, p_n)$ and $(b_{n,m}, p_m)$. Let 
\begin{talign}\label{FP_proof1}
    \mathit{J}(\vect{b},\vect{p_u},\vect{p_s}) =& {\mathit{M}_1 (p_n)}^2 z_{1,n,m}+\frac{1}{4 {\mathit{N}_1 (b_{n,m}, p_n)}^2 z_{1,n,m}} +\nonumber\\& {\mathit{M}_2 (p_m)}^2 z_{2,n,m}  + \frac{1}{4 {\mathit{N}_2 (b_{n,m}, p_m)}^2 z_{2,n,m}}.
\end{talign}
$\frac{p_n x_{n,m}\varphi_n d_n}{r_{n,m}} + \frac{p_m x_{n,m} \varphi_n d_n \omega_p}{r_{m,n}}$ can be expressed as 
\begin{talign}\label{FP_proof2}
    \frac{\mathit{M}_1(p_n)}{\mathit{N}_1 (b_{n,m}, p_n)} + \frac{\mathit{M}_2(p_m)}{\mathit{N}_2 (b_{n,m}, p_m)}.
\end{talign}
See the partial derivative of $\vect{b}$ of (\ref{FP_proof1}) and (\ref{FP_proof2}).
\begin{talign}\label{FP_proof3}
    &\frac{\partial (\frac{\mathit{M}_1(p_n)}{\mathit{N}_1 (b_{n,m}, p_n)} + \frac{\mathit{M}_2(p_m)}{\mathit{N}_2 (b_{n,m}, p_m)})}{\partial \vect{b}} = \nonumber \\ &- \frac{\mathit{M}_1(p_n)}{{\mathit{N}_1 (b_{n,m}, p_n)}^2} \frac{\partial \mathit{N}_1 (b_{n,m}, p_n)}{\partial \vect{b}} - \frac{\mathit{M}_2(p_m)}{{\mathit{N}_2 (b_{n,m}, p_m)}^2} \frac{\partial \mathit{N}_2 (b_{n,m}, p_m)}{\partial \vect{b}}.
\end{talign}
\begin{talign}\label{FP_proof4}
    \frac{\partial \mathit{J}(\vect{b},\vect{p_u},\vect{p_s})}{\partial \vect{b}} = - \frac{1}{2 z_{1,n,m} {\mathit{N}_1 (b_{n,m}, p_n)}^3}\frac{\partial \mathit{N}_1 (b_{n,m}, p_n)}{\partial \vect{b}} - &\nonumber \\ \frac{1}{2 z_{2,n,m} {\mathit{N}_2 (b_{n,m}, p_m)}^3}\frac{\partial \mathit{N}_2 (b_{n,m}, p_m)}{\partial \vect{b}}.&
\end{talign}
From Eq. (\ref{FP_proof3}) and Eq. (\ref{FP_proof4}), we can know that 
\begin{talign}
     (\frac{\partial \mathit{J}(\vect{b},\vect{p_u},\vect{p_s})}{\partial \vect{b}})&|_{z_{1,n,m}=\frac{1}{2 \mathit{M}_1(p_n)\mathit{N}_1 (b_{n,m}, p_n)},}\nonumber \\&_{z_{2,n,m}=\frac{1}{2 \mathit{M}_2(p_m)\mathit{N}_2 (b_{n,m}, p_m)}} \nonumber \\ &= \frac{\partial (\frac{\mathit{M}_1(p_n)}{\mathit{N}_1 (b_{n,m}, p_n)} + \frac{\mathit{M}_2(p_m)}{\mathit{N}_2 (b_{n,m}, p_m)})}{\partial \vect{b}}.
\end{talign}
By repeating the same steps mentioned above, we can obtain
\begin{talign}
    (\frac{\partial \mathit{J}(\vect{b},\vect{p_u},\vect{p_s})}{\partial \vect{p_u}})&|_{z_{1,n,m}=\frac{1}{2 \mathit{M}_1(p_n)\mathit{N}_1 (b_{n,m}, p_n)},}\nonumber \\&_{z_{2,n,m}=\frac{1}{2 \mathit{M}_2(p_m)\mathit{N}_2 (b_{n,m}, p_m)}} \nonumber\\&= \frac{\partial (\frac{\mathit{M}_1(p_n)}{\mathit{N}_1 (b_{n,m}, p_n)} + \frac{\mathit{M}_2(p_m)}{\mathit{N}_2 (b_{n,m}, p_m)})}{\partial \vect{p_u}},
\end{talign}
\begin{talign}
    (\frac{\partial \mathit{J}(\vect{b},\vect{p_u},\vect{p_s})}{\partial \vect{p_s}})&|_{z_{1,n,m}=\frac{1}{2 \mathit{M}_1(p_n)\mathit{N}_1 (b_{n,m}, p_n)},}\nonumber \\&_{z_{2,n,m}=\frac{1}{2 \mathit{M}_2(p_m)\mathit{N}_2 (b_{n,m}, p_m)}} \nonumber\\&= \frac{\partial (\frac{\mathit{M}_1(p_n)}{\mathit{N}_1 (b_{n,m}, p_n)} + \frac{\mathit{M}_2(p_m)}{\mathit{N}_2 (b_{n,m}, p_m)})}{\partial \vect{p_s}}.
\end{talign}
Furthermore, we can know that $\frac{\partial Eq. (\ref{fp_ori_prob})}{\partial (\vect{b},\vect{p_u},\vect{p_s})} = \frac{\partial Eq. (\ref{fp_trans_prob})}{\partial (\vect{b},\vect{p_u},\vect{p_s})}$ and Eq. (\ref{fp_ori_prob}) = Eq. (\ref{fp_trans_prob}) when $z_{1,n,m}=\frac{1}{2 \mathit{M}_1(p_n)\mathit{N}_1 (b_{n,m}, p_n)}$, $z_{2,n,m}=\frac{1}{2 \mathit{M}_2(p_m)\mathit{N}_2 (b_{n,m}, p_m)}$. The value of the AO procedure of Problem (\ref{prob10}) is non-decreasing, which means that at $i$ and $i+1$ iterations, 
\begin{talign}
    &V_{\text{AO-Part2}}(\vect{b}^{(i+1)}\!,\!\vect{f_s}^{(i+1)}\!,\!\vect{f_u}^{(i+1)}\!,\!\vect{p_s}^{(i+1)}\!,\!\vect{p_u}^{(i+1)}\!,\!T^{(i+1)}\!,\!\vect{z}_1^{(i+1)}\!,\!\vect{z}_2^{(i+1)}) \nonumber \\ &\geq  V_{\text{AO-Part2}}(\vect{b}^{(i)},\vect{f_s}^{(i)},\vect{f_u}^{(i)},\vect{p_s}^{(i)},\vect{p_u}^{(i)},T^{(i)},\vect{z}_1^{(i+1)},\vect{z}_2^{(i+1)}) \nonumber \\& \geq V_{\text{AO-Part2}}(\vect{b}^{(i)},\vect{f_s}^{(i)},\vect{f_u}^{(i)},\vect{p_s}^{(i)},\vect{p_u}^{(i)},T^{(i)},\vect{z}_1^{(i)},\vect{z}_2^{(i)}).
\end{talign}
As $i \!\!\rightarrow\!\! \infty$, $V_{\text{AO-Part2}}(\vect{b}^{(i)}\!,\vect{f_s}^{(i)}\!,\vect{f_u}^{(i)}\!,\vect{p_s}^{(i)}\!,\vect{p_u}^{(i)}\!,T^{(i)}\!,\vect{z}_1^{(i)}\!,\vect{z}_2^{(i)})$ converges to $V_{\text{AO-Part2}}(\vect{b}^{(\star)}\!,\!\vect{f_s}^{(\star)}\!,\!\vect{f_u}^{(\star)}\!,\!\vect{p_s}^{(\star)}\!,\!\vect{p_u}^{(\star)}\!,\!T^{(\star)}\!,\!\vect{z}_1^{(\star)}\!,\!\vect{z}_2^{(\star)}\!)$. With $\vect{z}_1^\star\!\!=\!\!\frac{1}{2 \mathit{M}_1(\vect{p_u}^\star)\mathit{N}_1 (\vect{b}^\star,\vect{p_u}^\star)}$ and $\vect{z}_2^\star\!\!=\!\!\frac{1}{2 \mathit{M}_2(\vect{p_s}^\star)\mathit{N}_2 (\vect{b}^\star,\vect{p_s}^\star)}$, the solution of (\ref{prob11}) is $(\vect{b}^{(\star)},\vect{f_s}^{(\star)},\vect{f_u}^{(\star)},\vect{p_s}^{(\star)},\vect{p_u}^{(\star)},T^{(\star)})$, which means that $(\vect{b}^{(\star)},\vect{f_s}^{(\star)},\vect{f_u}^{(\star)},\vect{p_s}^{(\star)},\vect{p_u}^{(\star)},T^{(\star)})$ is a stationary point of (\ref{prob10}). 
\begin{algorithm}
\caption{AO-Part2: Optimizing $\vect{b}$, $\vect{p_u}$, $\vect{p_s}$, $\vect{f_u}$, $\vect{f_s}$, $T$, Given $\vect{w}$, $\vect{\varphi}$, $\vect{\gamma}$.}
\label{algo:AO-part2}

Initialize $i \leftarrow -1$ and for all $n \in \mathcal{N}, m \in \mathcal{M}$: $b_{n,m}^{(0)} = \frac{b_{\max}}{N}$, $p_n^{(0)}=p_n$, $p_m^{(0)} = \frac{p_{\max}^{(m)}}{N}$, $F_n^{(0)} = F_{max}^{(n)}$, $F_{n,m}^{(0)} = \frac{F_{n,m}^{(m)}}{N}$;

$T^{(0)}\leftarrow \max_{n \in \mathcal{N}, m \in \mathcal{M}}(T_{u,n,m}^{(0)},T_{serever}^{(0)})$;

$y^{(0)} \leftarrow \sum\limits_{n \in \mathcal{N}, m \in \mathcal{M}} \frac{x_{n,m}^{(0)} v_{n,m}}{\omega_t T^{(0)} + \omega_e (E_{u,n,m}^{(0)} + E_{s,n,m}^{(0)})}$;

$z_{1,n,m}^{(0)} \leftarrow \frac{1}{2 p_n^{(0)} b_{n,m}^{(0)} \log_2(1+\frac{g_{n,m}p_n^{(0)}}{\delta^2 b_{n,m}^{(0)}})}$;

$z_{2,n,m}^{(0)} \leftarrow \frac{1}{2 p_m^{(0)} b_{n,m}^{(0)} \log_2(1+\frac{g_{n,m}p_m^{(0)}}{\delta^2 b_{n,m}^{(0)}})}$;

\Repeat{the relative difference between $V_{\text{AO-P2}}(\vect{b}^{(i+1)}, \vect{p_u}^{(i+1)}, \vect{p_s}^{(i+1)}, \vect{f_u}^{(i+1)}, \vect{f_s}^{(i+1)})$ and $V_{\text{AO-P2}}(\vect{b}^{(i)}, \vect{p_u}^{(i)}, \vect{p_s}^{(i)}, \vect{f_u}^{(i)}, \vect{f_s}^{(i)})$ is no greater than $\epsilon$ for a small positive number $\epsilon$ (i.e., $\frac{V_{\text{AO-P2}}(\vect{b}^{(i+1)}, \vect{p_u}^{(i+1)}, \vect{p_s}^{(i+1)}, \vect{f_u}^{(i+1)}, \vect{f_s}^{(i+1)})}{V_{\text{AO-P2}}(\vect{b}^{(i)}, \vect{p_u}^{(i)}, \vect{p_s}^{(i)}, \vect{f_u}^{(i)}, \vect{f_s}^{(i)})}- 1 \leq \epsilon$)}{

Let $i \leftarrow i+1$;

Obtain $[\vect{b}^{(i+1)}, \vect{p_u}^{(i+1)}, \vect{p_s}^{(i+1)}, \vect{f_u}^{(i+1)}, \vect{f_s}^{(i+1)}, T^{(i+1)}]$ by solving Problem (\ref{prob11}) with $[y^{(i)}, z_{1,n,m}^{(i)}, z_{2,n,m}^{(i)}]$;

Update $y^{(i+1)}$, $z_{1,n,m}^{(i+1)}$, and $z_{2,n,m}^{(i+1)}$ with the solution $[\vect{b}^{(i+1)}, \vect{p_u}^{(i+1)}, \vect{p_s}^{(i+1)}, \vect{f_u}^{(i+1)}, \vect{f_s}^{(i+1)}, T^{(i+1)}]$;

$y^{(i+1)} \leftarrow \sum\limits_{n \in \mathcal{N}, m \in \mathcal{M}} \frac{x_{n,m}^{(i+1)} v_{n,m}}{\omega_t T^{(i+1)} + \omega_e (E_{u,n,m}^{(i+1)} + E_{s,n,m}^{(i+1)})}$;

$z_{1,n,m}^{(i+1)} \leftarrow \frac{1}{2 p_n^{(i+1)} b_{n,m}^{(i+1)} \log_2(1+\frac{g_{n,m}p_n^{(i+1)}}{\delta^2 b_{n,m}^{(i+1)}})}$;

$z_{2,n,m}^{(i+1)} \leftarrow \frac{1}{2 p_m^{(i+1)} b_{n,m}^{(i+1)} \log_2(1+\frac{g_{n,m}p_m^{(i+1)}}{\delta^2 b_{n,m}^{(i+1)}})}$;

}

Return $[\boldsymbol{b}^{(i+1)},\boldsymbol{p_u}^{(i+1)},\boldsymbol{p_s}^{(i+1)},\boldsymbol{f_u}^{(i+1)},\boldsymbol{f_s}^{(i+1)}]$ as a solution $[\boldsymbol{b}^{\star},\boldsymbol{p_u}^{\star},\boldsymbol{p_s}^{\star},\boldsymbol{f_u}^{\star},\boldsymbol{f_s}^{\star}]$ to Problem $\text{AO-Part2}$.
\end{algorithm}

\begin{algorithm}
\caption{The Whole Process of The DASHF Algorithm.}
\label{algo:AO-p3}

Initialize $i \leftarrow -1$ and for all $n \in \mathcal{N}, m \in \mathcal{M}$: $b_n^{(0)} = \frac{b_{\max}}{N}$, $p_n^{(0)}=p_n$, $p_m^{(0)} = \frac{p_{\max}^{(m)}}{N}$, $F_n^{(0)} = F_{max}^{(n)}$, $F_{n,m}^{(0)} = \frac{F_{n,m}^{(m)}}{N}$, $\vect{x}^{(0)} = (\vect{e_1},\cdots,\vect{e_M})^\intercal$, $\varphi_n^{(0)}=0.5$, $\gamma_{n,m} = \frac{\omega_b}{\omega_b + 1}$;

$y^{(0)} \leftarrow \sum\limits_{n \in \mathcal{N}, m \in \mathcal{M}} \frac{x_{n,m}^{(0)} v_{n,m}}{\omega_t T^{(0)} + \omega_e (E_{u,n,m}^{(0)} + E_{s,n,m}^{(0)})}$;

\Repeat{the relative difference between $y^{(i+1)}$ and $y^{(i)}$ is no greater than $\epsilon$ for a small positive number $\epsilon$ (i.e., $\frac{y^{(i+1)}}{y^{(i)}}- 1 \leq \epsilon$)}{

Let $i \leftarrow i+1$;

Obtain $[\vect{x}^{(i+1)},\vect{\varphi}^{(i+1)}]$ by using \textbf{Algorithm \ref{algo:AO-part1}};

Obtain $[\boldsymbol{b}^{(i+1)},\boldsymbol{p_u}^{(i+1)},\boldsymbol{p_s}^{(i+1)},\boldsymbol{f_u}^{(i+1)},\boldsymbol{f_s}^{(i+1)}]$ by using \textbf{Algorithm \ref{algo:AO-part2}};

Update $y^{(i+1)} \leftarrow \sum\limits_{n \in \mathcal{N}, m \in \mathcal{M}} \frac{x_{n,m}^{(i+1)} v_{n,m}}{\omega_t T^{(i+1)} + \omega_e (E_{u,n,m}^{(i+1)} + E_{s,n,m}^{(i+1)})}$ with $[\vect{x}^{(i+1)},\vect{\varphi}^{(i+1)},\boldsymbol{b}^{(i+1)},\boldsymbol{p_u}^{(i+1)},\boldsymbol{p_s}^{(i+1)},\boldsymbol{f_u}^{(i+1)},\boldsymbol{f_s}^{(i+1)}]$;
}

Set $[\vect{x}^{(i+1)},\vect{\varphi}^{(i+1)},\boldsymbol{b}^{(i+1)},\boldsymbol{p_u}^{(i+1)},\boldsymbol{p_s}^{(i+1)},\boldsymbol{f_u}^{(i+1)},\boldsymbol{f_s}^{(i+1)}]$ as a solution to Problem (\ref{prob1});

Return $[\vect{x}^{(i+1)},\vect{\varphi}^{(i+1)},\boldsymbol{b}^{(i+1)},\boldsymbol{p_u}^{(i+1)},\boldsymbol{p_s}^{(i+1)},\boldsymbol{f_u}^{(i+1)},\boldsymbol{f_s}^{(i+1)}]$ as a solution $[\vect{x}^{\star},\vect{\varphi}^{\star},\boldsymbol{b}^{\star},\boldsymbol{p_u}^{\star},\boldsymbol{p_s}^{\star},\boldsymbol{f_u}^{\star},\boldsymbol{f_s}^{\star}]$ to Problem~(\ref{prob1}).
\end{algorithm}

\subsection{Complexity Analysis of the Whole Algorithm}\label{section.Complexity Analysis}
In Algorithm \ref{algo:AO-part1}, at each iteration, there are $a_1 = N(M+2)+1$ variables and $b_1 = N(4M+4)+3$ constraints. The complexity of the Hungarian algorithm is $\mathcal{O}(N^3)$. The worst-case complexity of Algorithm \ref{algo:AO-part1} is $\mathcal{O}(N^3+(a_1^2b_1+a_1^3)b_1^{0.5}\log(\frac{1}{\epsilon_1}))$ with a given solution accuracy $\epsilon_1 > 0$ \cite{dai2018joint}. In Algorithm \ref{algo:AO-part2}, at each iteration, there are $a_2 = N(3M+2)+1$ variables and $b_2 = N(5M+4)+4$ constraints. The worst-case complexity of Algorithm \ref{algo:AO-part2} is $\mathcal{O}(a_2^2b_2+a_2^3)b_2^{0.5}\log(\frac{1}{\epsilon_2})$ with a given solution accuracy $\epsilon_2 > 0$. To summarize, if Algorithm \ref{algo:AO-p3} takes $\mathcal{I}$ iterations, the whole complexity is $\mathcal{I}\mathcal{O}(N^3+(a_1^2b_1+a_1^3)b_1^{0.5}\log(\frac{1}{\epsilon_1}) + (a_2^2b_2+a_2^3)b_2^{0.5}\log(\frac{1}{\epsilon_2}))$.

\section{Numerical Results}\label{section.Numerical Results}
In this section, we first introduce the default settings for the numerical simulations. Subsequently, we verify the convergence of the proposed DASHF algorithm and compare it with other baselines to validate its effectiveness. We adjust the available communication and computational resources and the cost weight parameters to analyze their impacts on the TCR.
\subsection{Default settings}
\begin{figure}[tbp]
\centering
\includegraphics[width=0.37\textwidth]{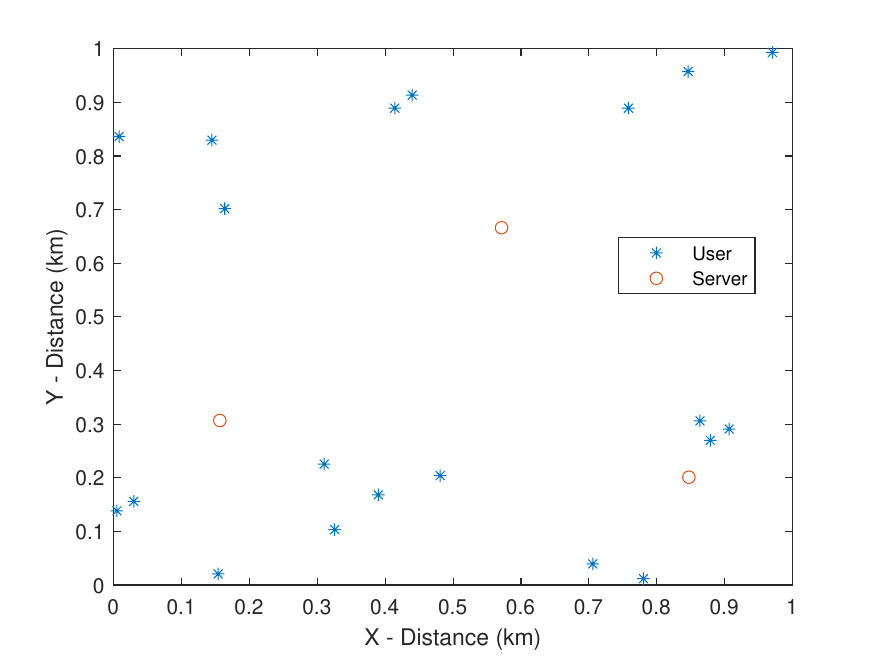}
\vspace{-6pt}\caption{2D distribution of 3 servers and 20 users.}
\label{Fig.2D distribution of server and user}
\end{figure}

We consider a network topology of 1000 m $\times$ 1000 m with 20 mobile users and three servers in Fig. \ref{Fig.2D distribution of server and user}. The large-scale fading $h_{n,m}$ between the user $n$ and server $m$ is modeled as $128.1+37.6 \log_{10}d_{n,m}$\cite{dai2018joint}, where $d_{n,m}$ denotes the  Euclidean distance between the user $n$ and server $m$. The small-scale fading is the Rayleigh fading. Gaussian noise power $\sigma^2$ is $-134$dBm. The total bandwidth for each server $b_{max}$ is 10 MHz. The maximum transmit power of mobile users $p_{max}^{(n)}$ is 0.2 W. The maximum transmit power of servers $p_{max}^{(m)}$ is 10 W. The maximum computation capacity of mobile users $F_{max}^{(n)}$ is 1 GHz, and that of servers $F_{max}^{(m)}$ is 20 GHz. The CPU clock cycles per bit for processing data of mobile users and servers ($f_n$ and $f_m$) are 279.62 \cite{feng2020joint}. The CPU clock cycles per bit for generating blocks of servers ($f_m$ at the blockchain phase) are 737.5 \cite{feng2020joint}. The effective switched capacitance of mobile users and servers ($\kappa_n$ and $\kappa_m$) is $10^{-27}$ to be the energy consumption in a reasonable range. The data size change ratio between the raw data and processed data $\omega_p$ is 0.9. The data sizes of mobile users are randomly selected from $[500\text{KB}, 2000 \text{KB}]$. To achieve this, we generate pseudorandom values, which follow a standard uniform distribution over the open interval $(0,1)$. These pseudorandom values are then scaled to the range of $[500\text{KB}, 2000 \text{KB}]$ to determine the specific data sizes for each mobile user.
The block size $S_b$ is 8 MB. The data rate of wired links among servers $R_{m}$ is 15 Mbps (M bits/second). The data size change ratio between the MEC task and the blockchain task $\omega_b$ is 1. The parameters of delay and energy consumption ($\omega_t$ and $\omega_e$) are 0.5 and 0.5. We set $\varpi_1$ and $\varpi_2$ as $\frac{100}{\ln2}$ and $0.25$, respectively. We consider using the Mosek optimization tool in Matlab to conduct the simulations.
\begin{figure*}[t]
\vspace{-0.9cm}
\subfigure[Convergence of Algorithm AO-Part 1.]{\includegraphics[width=.3\textwidth]{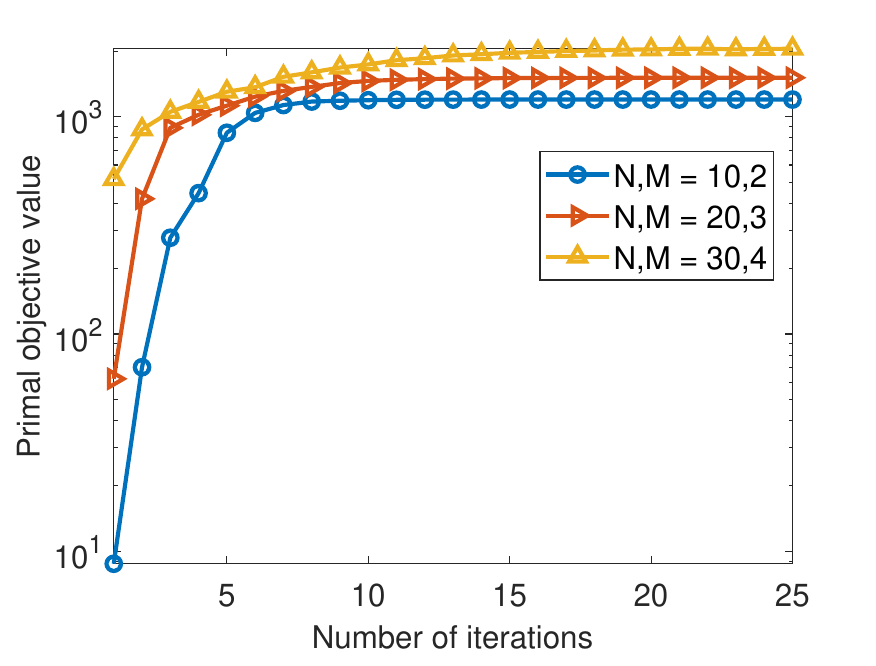}\vspace{-10pt}\label{fig:Convergence_of_AO-Part1}}
\subfigure[Convergence of Algorithm AO-Part 2.]{\includegraphics[width=.3\textwidth]{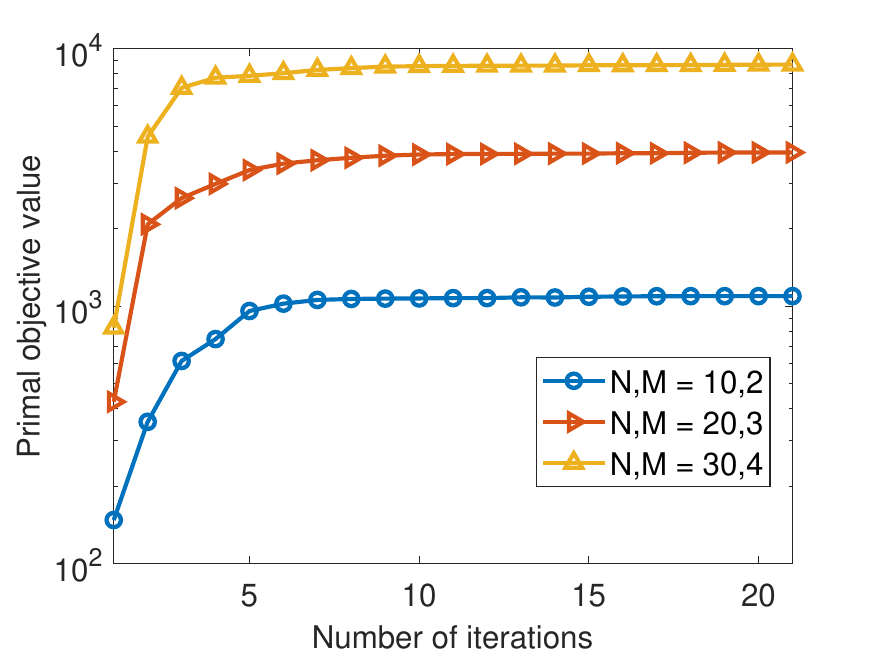}\vspace{-10pt}\label{fig:Convergence_of_AO-Part2}}
\subfigure[Convergence of Algorithm DASHF.]{\includegraphics[width=.3\textwidth]{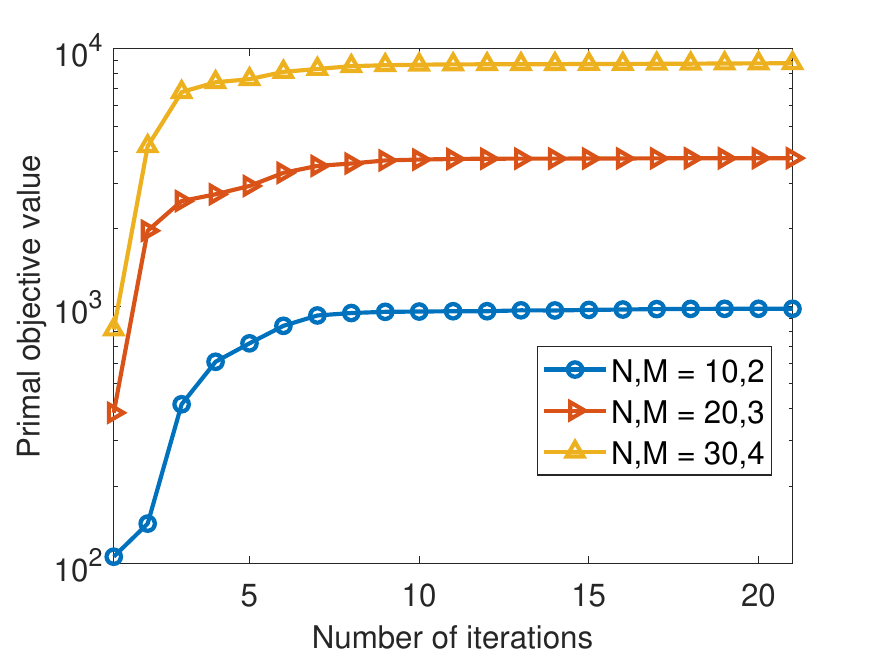}\vspace{-10pt}\label{fig:Convergence_of_AO}}
\vspace{-5pt}\caption{Convergence of the proposed Algorithms.}
\end{figure*}
\subsection{Convergence of proposed Algorithms}
In this section, we evaluate the convergence of the proposed algorithms. We consider three network topologies with $(N,M)=(10,2),(20,3),(30,4)$ and keep other settings as default. Primal objective value means an estimate for the primal objective value when using Mosek to solve the optimization problem. When the primal objective value converges, the algorithm converges to one stationary point. Fig. \ref{fig:Convergence_of_AO-Part1} plots the convergence of the Algorithm AO-Part 1, which converges within 15 iterations. Fig. \ref{fig:Convergence_of_AO-Part2} plots the convergence of the Algorithm AO-Part 2, which converges within nine iterations. Fig. \ref{fig:Convergence_of_AO} plots the convergence of the DASHF Algorithm, which converges within nine iterations. Thus, the proposed DASHF Algorithm is effective in finding one stationary point of the Problem (\ref{prob1}). In Table \ref{tab:DASHF_runningtime}, we have presented the running time of the proposed DASHF algorithm for those three network topologies.
\begin{table}[!htbp]
\centering
\caption{Running time of the proposed DASHF Algorithm}
\label{tab:DASHF_runningtime}
\begin{tabularx}{0.46\textwidth}{@{}Xccc@{}}
\toprule
\textbf{N, M} & \textbf{10, 2} & \textbf{20, 3} & \textbf{30, 4} \\
\midrule
time consumed & 3.07s & 26.94s & 1600.09s \\
\bottomrule
\end{tabularx}
\end{table}
\subsection{Comparison with baselines}
\begin{figure}[tbp]
\centering
\vspace{-15pt}\includegraphics[width=0.38\textwidth]{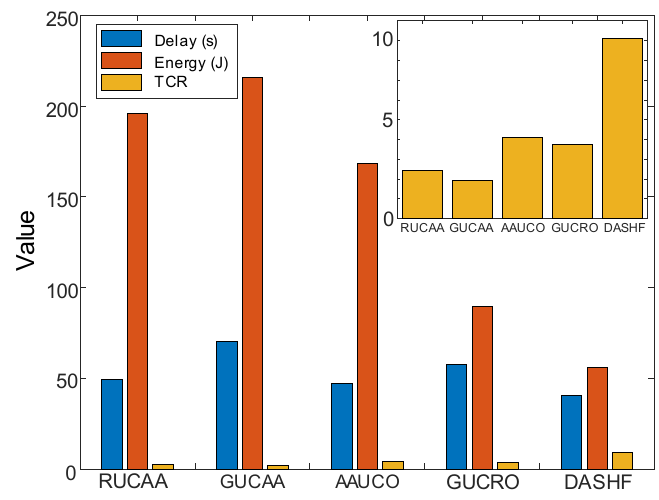}
\vspace{-2pt}\caption{Resource consumption and TCR of different baselines and proposed AO Algorithm.}
\label{Fig.AO performance}
\end{figure}

In this section, we mainly consider four baselines to carry out the comparison experiments.
\begin{enumerate}
    \item \textbf{\underline{R}andom \underline{u}ser \underline{c}onnection with \underline{a}verage resource \underline{a}llocation (RUCAA)}. In this algorithm, one server is randomly selected for each user. The server equally allocates communication and computational resources among the users connected to it.
    \item \textbf{\underline{G}reedy \underline{u}ser \underline{c}onnection with \underline{a}verage resource \underline{a}llocation (GUCAA)}. In this algorithm, each user selects the server with the least number of users underserving. The server distributes communication and computational resources evenly to the users connected to it.
    \item \textbf{\underline{A}verage resource \underline{a}llocation with \underline{u}ser \underline{c}onnection \underline{o}ptimization (AAUCO)}. In this algorithm, the communication and computation resources of each Metaverse server are equally allocated to each user connected to it. Besides, \textbf{Algorithm \ref{algo:AO-part1}} is leveraged to operate user connection optimization.
    \item \textbf{\underline{G}reedy \underline{u}ser \underline{c}onnection with \underline{r}esource allocation \underline{o}ptimization (GUCRO)}. In this algorithm, each user selects the Metaverse server with the least number of users underserving. Besides, \textbf{Algorithm \ref{algo:AO-part2}} is leveraged to operate resource optimization.
    \item \textbf{Proposed DASHF algorithm}. Joint optimization of user connection and resource allocation by utilizing \textbf{Algorithm \ref{algo:AO-p3}}.
\end{enumerate}
In Fig. \ref{Fig.AO performance}, we compare the resource consumption and TCR of the proposed DASHF Algorithms with other baselines. The performances of RUCAA and GUCAA are worse since no optimization is utilized. GUCRO and AAUCO have better performances than RUCAA and GUCAA, which confirms the effectiveness of the proposed Algorithm AO-Part 1 and Part 2. Furthermore, the TCR of GUCRO is higher than that of AAUCO, which shows that resource optimization is more effective than user connection optimization. The resource consumption of the proposed DASHF algorithm is the lowest of these five methods, and the TCR is also the highest one. This results from the benefits of joint optimization of user connection and resource allocation.
\subsection{TCR versus resources}
For communication and computation resources comparison experiments, we consider the total bandwidth and the computing frequency of servers, which are the dominant factors.
\begin{enumerate}
    \item TCR versus the total bandwidth. We consider the total bandwidth from 10 MHz to 100 MHz to test the TCR under different total bandwidths. Other parameters are fixed as default settings. Fig. \ref{Fig.Comparison_of_b_max} reveals distinct algorithmic performance trends, with the proposed DASHF method consistently outperforming GUCRO, AAUCO, RUCAA, and GUCAA in terms of the TCR. Notably, optimization algorithms (GUCRO and AAUCO) demonstrate superior performance compared to non-optimization algorithms (RUCAA and GUCAA). While AAUCO employs user connection optimization strategies and performs better than RUCAA and GUCAA, it lags behind the resource optimization algorithms. 
    \begin{figure}[t]
    \vspace{-5pt}
    \centering\includegraphics[width=0.4\textwidth]{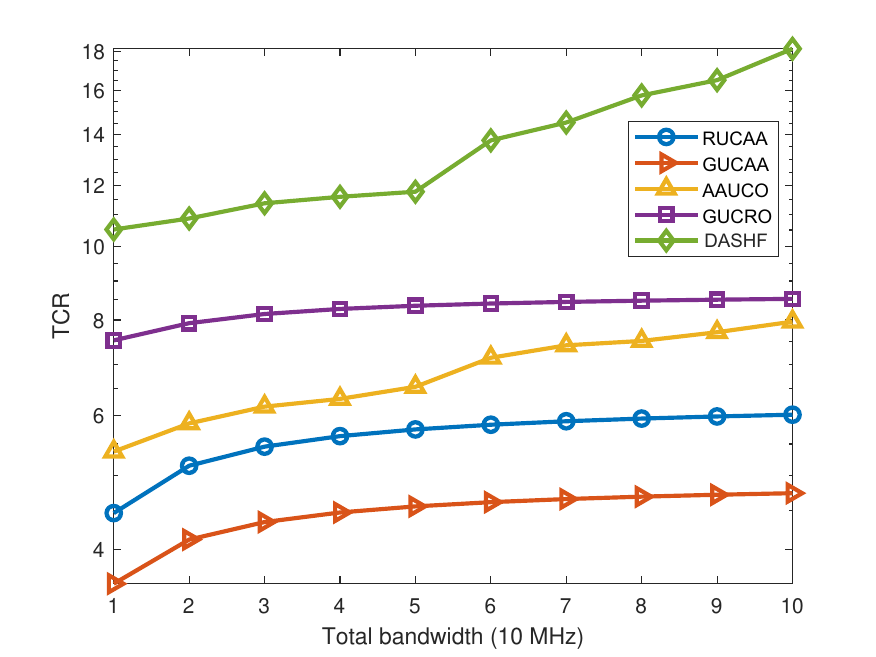}
    \vspace{-5pt}\caption{TCR under different total bandwidths.}
    \label{Fig.Comparison_of_b_max}
    \end{figure}
    
    \item TCR versus the computation resource. We consider changing the servers' computing frequencies from 20 GHz to 200 GHz to see the TCR trend under different servers' computing frequencies in Fig. \ref{Fig.Comparison_of_f_smax}. Other parameters are fixed as default settings. Fig. \ref{Fig.Comparison_of_f_smax} reveals an inverse relationship between the total computing frequency and algorithm performance, where increasing the total computing frequency leads to diminishing algorithm performance, as evidenced by decreasing TCR values. Importantly, this phenomenon can be attributed to the trade-off between computational resources and energy consumption. As servers' computing frequency increases with the expanded range, energy consumption rises correspondingly. Given that the TCR primarily reflects the ratio of computing frequency allocated to users relative to the total server computing frequency, it tends to have trivial impacts under these conditions. The proposed DASHF method consistently maintains its superior performance across the computing frequency range, while GUCRO demonstrates little change. Conversely, algorithms with no optimization, such as RUCAA and GUCAA, exhibit diminishing returns with computing resource expansion, emphasizing the significance of resource allocation strategies in computing resource-rich environments. AAUCO, though effective, falls short of the performance levels achieved by GUCRO and the proposed DASHF method.
    
    \begin{figure}[t]
    \centering
    \vspace{-25pt}\includegraphics[width=0.4\textwidth]{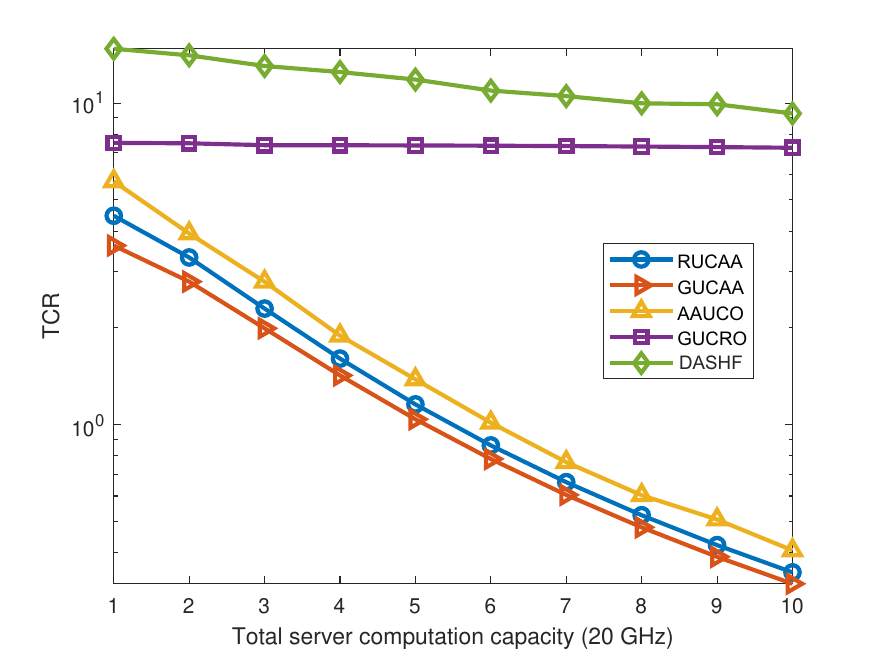}
    \vspace{-5pt}\caption{TCR under different server computation frequencies.}
    \label{Fig.Comparison_of_f_smax}
    \end{figure}
\end{enumerate}

\subsection{Impact of cost weights on TCR}
\begin{figure}[tbp]
\centering
\vspace{-10pt}\includegraphics[width=0.43\textwidth]{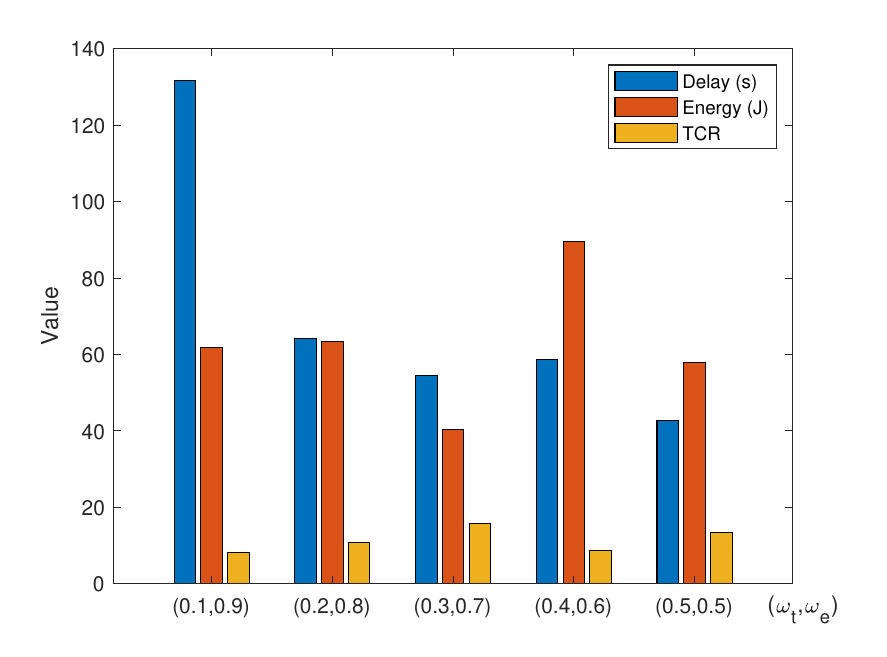}
\vspace{-12pt}\caption{Resources consumption and TCR under different $(\omega_{t},\omega_{e})$.}\vspace{-10pt}
\label{Fig.Comparison_of_omega}
\end{figure}

Fig. \ref{Fig.Comparison_of_omega} featuring various combinations of ($\omega_t$, $\omega_e$) that signifies the trade-off between delay-energy optimization and trust score. As ($\omega_t$, $\omega_e$) values shift, emphasizing either delay or energy, distinct performance outcomes are evident. For instance, when prioritizing energy efficiency (e.g., ($\omega_t$, $\omega_e$) = (0.1, 0.9)), the system achieves lower energy consumption but at the expense of higher delay, resulting in a moderate TCR value. Conversely, balanced settings (e.g., ($\omega_t$, $\omega_e$) = (0.5, 0.5)) lead to lower delay and slightly higher energy consumption, yielding a higher TCR value. These findings underscore the sensitivity of the optimization process to parameter choices and emphasize the importance of tailoring ($\omega_t$, $\omega_e$) values to meet specific application requirements while carefully considering the trade-offs between delay-energy and trust score.
\section{Conclusion}\label{section.Conclusion}
In this exploration of blockchain, the Metaverse, and NFT applications, we have navigated the complexities of optimizing user connection and resource allocation within 6G wireless communication systems. The Metaverse, once a concept of science fiction, has become a digital reality where NFTs play a transformative role. Our contributions, highlighted in this paper, pave the way for NFT integration into the Metaverse, offering solutions to intricate challenges. By introducing work offloading to servers, optimizing resource allocation, and introducing the trust-cost ratio, we enhance trust mechanisms and efficiency within this ecosystem. The proposed DASHF algorithm further streamlines these optimizations. As we conclude, we envision a future where the blockchain-empowered Metaverse seamlessly integrates NFTs, offering unprecedented digital experiences while upholding trust and resource efficiency.

\begin{spacing}{.98}
\bibliographystyle{IEEEtran}
\bibliography{ref}

\begin{thebibliography}{10}
\providecommand{\url}[1]{#1}
\csname url@samestyle\endcsname
\providecommand{\newblock}{\relax}
\providecommand{\bibinfo}[2]{#2}
\providecommand{\BIBentrySTDinterwordspacing}{\spaceskip=0pt\relax}
\providecommand{\BIBentryALTinterwordstretchfactor}{4}
\providecommand{\BIBentryALTinterwordspacing}{\spaceskip=\fontdimen2\font plus
\BIBentryALTinterwordstretchfactor\fontdimen3\font minus
  \fontdimen4\font\relax}
\providecommand{\BIBforeignlanguage}[2]{{%
\expandafter\ifx\csname l@#1\endcsname\relax
\typeout{** WARNING: IEEEtran.bst: No hyphenation pattern has been}%
\typeout{** loaded for the language `#1'. Using the pattern for}%
\typeout{** the default language instead.}%
\else
\language=\csname l@#1\endcsname
\fi
#2}}
\providecommand{\BIBdecl}{\relax}
\BIBdecl

\bibitem{qian2023user}
\BIBentryALTinterwordspacing
L.~Qian and J.~Zhao, ``User association and resource allocation in large
  language model based mobile edge computing system over wireless
  communications,'' \emph{accepted by IEEE Vehicular Technology Conference
  (VTC)}, 2024. [Online]. Available: \url{https://arxiv.org/pdf/2310.17872.pdf}
\BIBentrySTDinterwordspacing

\bibitem{zhao2023human}
J.~Zhao, L.~Qian, and W.~Yu, ``Human-centric resource allocation in the
  {Metaverse} over wireless communications,'' \emph{IEEE Journal on Selected
  Areas in Communications (JSAC)}, vol.~42, no.~3, pp. 514--537, 2024.

\bibitem{wang2022toward}
C.~Wang, C.~Yu, and Y.~Li, ``Toward understanding attention economy in
  {Metaverse}: A case study of {NFT} value,'' \emph{IEEE Transactions on
  Computational Social Systems}, 2022.

\bibitem{lim2022realizing}
W.~Y.~B. Lim, Z.~Xiong, D.~Niyato, X.~Cao, C.~Miao, S.~Sun, and Q.~Yang,
  ``Realizing the {Metaverse} with edge intelligence: A match made in heaven,''
  \emph{IEEE Wireless Communications}, 2022.

\bibitem{xu2022full}
M.~Xu, W.~C. Ng, W.~Y.~B. Lim, J.~Kang, Z.~Xiong, D.~Niyato, Q.~Yang, X.~S.
  Shen, and C.~Miao, ``A full dive into realizing the edge-enabled metaverse:
  Visions, enabling technologies, and challenges,'' \emph{IEEE Communications
  Surveys \& Tutorials}, 2022.

\bibitem{xu2023epvisa}
M.~Xu, D.~Niyato, B.~Wright, H.~Zhang, J.~Kang, Z.~Xiong, S.~Mao, and Z.~Han,
  ``Epvisa: Efficient auction design for real-time physical-virtual
  synchronization in the human-centric metaverse,'' \emph{IEEE Journal on
  Selected Areas in Communications}, 2023.

\bibitem{wu2023virtual}
D.~Wu, Z.~Yang, P.~Zhang, R.~Wang, B.~Yang, and X.~Ma, ``Virtual-reality
  inter-promotion technology for {Metaverse}: A survey,'' \emph{IEEE Internet
  of Things Journal}, 2023.

\bibitem{wang2022survey}
Y.~Wang, Z.~Su, N.~Zhang, R.~Xing, D.~Liu, T.~H. Luan, and X.~Shen, ``A survey
  on {Metaverse}: Fundamentals, security, and privacy,'' \emph{IEEE
  Communications Surveys \& Tutorials}, 2022.

\bibitem{christodoulou2022nfts}
K.~Christodoulou, L.~Katelaris, M.~Themistocleous, P.~Christodoulou, and
  E.~Iosif, ``{NFTs} and the {M}etaverse revolution: Research perspectives and
  open challenges,'' \emph{Blockchains and the Token Economy: Theory and
  Practice}, pp. 139--178, 2022.

\bibitem{chalmers2022beyond}
D.~Chalmers, C.~Fisch, R.~Matthews, W.~Quinn, and J.~Recker, ``Beyond the
  bubble: Will {NFTs} and digital proof of ownership empower creative industry
  entrepreneurs?'' \emph{Journal of Business Venturing Insights}, vol.~17, p.
  e00309, 2022.

\bibitem{cheng2022will}
R.~Cheng, N.~Wu, S.~Chen, and B.~Han, ``Will {Metaverse} be {NextG} {Internet}?
  {Vision}, hype, and reality,'' \emph{IEEE Network}, vol.~36, no.~5, pp.
  197--204, 2022.

\bibitem{aggarwal2019blockchain}
S.~Aggarwal, R.~Chaudhary, G.~S. Aujla, N.~Kumar, K.-K.~R. Choo, and A.~Y.
  Zomaya, ``Blockchain for smart communities: Applications, challenges and
  opportunities,'' \emph{Journal of Network and Computer Applications}, vol.
  144, pp. 13--48, 2019.

\bibitem{lu2019blockchain}
Y.~Lu, ``The blockchain: State-of-the-art and research challenges,''
  \emph{Journal of Industrial Information Integration}, vol.~15, pp. 80--90,
  2019.

\bibitem{yang2022fusing}
Q.~Yang, Y.~Zhao, H.~Huang, Z.~Xiong, J.~Kang, and Z.~Zheng, ``Fusing
  blockchain and {AI} with {Metaverse}: A survey,'' \emph{IEEE Open Journal of
  the Computer Society}, vol.~3, pp. 122--136, 2022.

\bibitem{huang2023security}
Y.~Huang, Y.~J. Li, and Z.~Cai, ``Security and privacy in {M}etaverse: A
  comprehensive survey,'' \emph{Big Data Mining and Analytics}, vol.~6, no.~2,
  pp. 234--247, 2023.

\bibitem{di2021metaverse}
R.~Di~Pietro and S.~Cresci, ``Metaverse: Security and privacy issues,'' in
  \emph{IEEE International Conference on Trust, Privacy and Security in
  Intelligent Systems and Applications}.\hskip 1em plus 0.5em minus 0.4em\relax
  IEEE, 2021, pp. 281--288.

\bibitem{kang2023security}
G.~Kang, J.~Koo, and Y.-G. Kim, ``Security and privacy requirements for the
  {M}etaverse: A {M}etaverse applications perspective,'' \emph{IEEE
  Communications Magazine}, 2023.

\bibitem{far2022applying}
S.~B. Far and A.~I. Rad, ``Applying digital twins in {M}etaverse: User
  interface, security and privacy challenges,'' \emph{Journal of Metaverse},
  vol.~2, no.~1, pp. 8--15, 2022.

\bibitem{feng2020joint}
J.~Feng, F.~R. Yu, Q.~Pei, J.~Du, and L.~Zhu, ``Joint optimization of radio and
  computational resources allocation in blockchain-enabled mobile edge
  computing systems,'' \emph{IEEE Transactions on Wireless Communications},
  vol.~19, no.~6, pp. 4321--4334, 2020.

\bibitem{dai2018joint}
Y.~Dai, D.~Xu, S.~Maharjan, and Y.~Zhang, ``Joint computation offloading and
  user association in multi-task mobile edge computing,'' \emph{IEEE
  Transactions on Vehicular Technology}, vol.~67, no.~12, pp. 12\,313--12\,325,
  2018.

\bibitem{lv2021strategy}
H.~Lv, Z.~Zheng, F.~Wu, and G.~Chen, ``Strategy-proof online mechanisms for
  weighted {AoI} minimization in edge computing,'' \emph{IEEE Journal on
  Selected Areas in Communications}, vol.~39, no.~5, pp. 1277--1292, 2021.

\bibitem{guo2019adaptive}
F.~Guo, F.~R. Yu, H.~Zhang, H.~Ji, M.~Liu, and V.~C. Leung, ``Adaptive resource
  allocation in future wireless networks with blockchain and mobile edge
  computing,'' \emph{IEEE Transactions on Wireless Communications}, vol.~19,
  no.~3, pp. 1689--1703, 2019.

\bibitem{li2024trajectory}
H.~Li, Y.~Huo, S.~Dou, Z.~Zheng, Z.~Zhang, C.~Yu, J.~Xu, and F.~Wu,
  ``Trajectory-wise iterative reinforcement learning framework for
  auto-bidding,'' in \emph{Proceedings of the ACM on Web Conference (WWW)},
  2024, pp. 4193--4203.

\bibitem{sun2020joint}
W.~Sun, J.~Liu, Y.~Yue, and P.~Wang, ``Joint resource allocation and incentive
  design for blockchain-based mobile edge computing,'' \emph{IEEE Transactions
  on Wireless Communications}, vol.~19, no.~9, pp. 6050--6064, 2020.

\bibitem{zheng2014star}
Z.~Zheng, Y.~Gui, F.~Wu, and G.~Chen, ``{STAR}: Strategy-proof double auctions
  for multi-cloud, multi-tenant bandwidth reservation,'' \emph{IEEE
  Transactions on Computers}, vol.~64, no.~7, pp. 2071--2083, 2014.

\bibitem{zheng2014unknown}
Z.~Zheng, F.~Wu, S.~Tang, and G.~Chen, ``Unknown combinatorial auction
  mechanisms for heterogeneous spectrum redistribution,'' in \emph{15th ACM
  International Symposium on Mobile Ad-Hoc Networking and Computing (ACM
  MobiHoc)}, 2014, pp. 3--12.

\bibitem{xu2021edge}
H.~Xu, W.~Huang, Y.~Zhou, D.~Yang, M.~Li, and Z.~Han, ``Edge computing resource
  allocation for unmanned aerial vehicle assisted mobile network with
  blockchain applications,'' \emph{IEEE Transactions on Wireless
  Communications}, vol.~20, no.~5, pp. 3107--3121, 2021.

\bibitem{jiang2020intelligent}
X.~Jiang, F.~R. Yu, T.~Song, and V.~C. Leung, ``Intelligent resource allocation
  for video analytics in blockchain-enabled {Internet} of autonomous vehicles
  with edge computing,'' \emph{IEEE Internet of Things Journal}, vol.~9,
  no.~16, pp. 14\,260--14\,272, 2020.

\bibitem{tu2022blockchain}
Z.~Tu, H.~Zhou, K.~Li, H.~Song, and Y.~Yang, ``A blockchain-based trust and
  reputation model with dynamic evaluation mechanism for {IoT},''
  \emph{Computer Networks}, vol. 218, p. 109404, 2022.

\bibitem{liu2022semi}
Y.~Liu, C.~Zhang, Y.~Yan, X.~Zhou, Z.~Tian, and J.~Zhang, ``A semi-centralized
  trust management model based on blockchain for data exchange in {IoT}
  system,'' \emph{IEEE Transactions on Services Computing}, vol.~16, no.~2, pp.
  858--871, 2022.

\bibitem{xi2023blockchain}
J.~Xi, G.~Xu, S.~Zou, Y.~Lu, G.~Li, J.~Xu, and R.~Wang, ``A blockchain dynamic
  sharding scheme based on hidden {M}arkov model in collaborative {IoT},''
  \emph{IEEE Internet of Things Journal}, 2023.

\bibitem{hoa2023dynamic}
N.~T. Hoa, B.~D. Son, N.~C. Luong, and D.~Niyato, ``Dynamic offloading for edge
  computing-assisted {Metaverse} systems,'' \emph{IEEE Communications Letters},
  2023.

\bibitem{monkeykingdom}
\BIBentryALTinterwordspacing
``Monkey {Kingdom}.'' [Online]. Available: \url{https://monkeykingdom.io/}
\BIBentrySTDinterwordspacing

\bibitem{yang2020energy}
Z.~Yang, M.~Chen, W.~Saad, C.~S. Hong, and M.~Shikh-Bahaei, ``Energy efficient
  federated learning over wireless communication networks,'' \emph{IEEE
  Transactions on Wireless Communications}, vol.~20, no.~3, pp. 1935--1949,
  2020.

\bibitem{al2021blockchain}
M.~S. Al-Rakhami and M.~Al-Mashari, ``A blockchain-based trust model for the
  {Internet} of {Things} supply chain management,'' \emph{Sensors}, vol.~21,
  no.~5, p. 1759, 2021.

\bibitem{yang2015incentive}
D.~Yang, G.~Xue, X.~Fang, and J.~Tang, ``Incentive mechanisms for crowdsensing:
  Crowdsourcing with smartphones,'' \emph{IEEE/ACM Transactions on Networking},
  vol.~24, no.~3, pp. 1732--1744, 2015.

\bibitem{dinkelbach1967nonlinear}
W.~Dinkelbach, ``On nonlinear fractional programming,'' \emph{Management
  Science}, vol.~13, no.~7, pp. 492--498, 1967.

\end{thebibliography}
\end{spacing}
\end{document}